\documentclass[aps,prb,twocolumn]{revtex4} 

\usepackage{graphicx}
\usepackage{dcolumn}
\usepackage{bm}

\newcommand{\comment}[1]{}

\begin{document}
\renewcommand{\theequation}{\arabic{section}.\arabic{equation}}

\title{Quantum Statistical Mechanics. III. Equilibrium Probability}


\author{Phil Attard}

\date{Apr.\ 3, 2014. phil.attard1@gmail.com}

\begin{abstract}
Given are a first principles derivation and formulation
of the probabilistic concepts that underly
equilibrium quantum statistical mechanics.
The transition to non-equilibrium probability is  traversed briefly.
\end{abstract}


\maketitle

\section*{Introduction}
\setcounter{equation}{0}

This paper continues a series that ultimately aims to establish
the theory of non-equilibrium quantum statistical mechanics.
The first two papers in the series dealt with the canonical
equilibrium system, namely a sub-system able to exchange energy
with a fixed temperature reservoir.\cite{QSM1,QSM2}

The first paper\cite{QSM1} showed that the wave function collapsed
into entropy pure quantum states,
due to entanglement resulting from energy conservation,
and that consequently the average of an operator observable
could be written as the sum over quantum states
of the product of the probability and observable operators.
The probability operator was shown to be the usual
Maxwell-Boltzmann operator.
These three results ---the wave function collapse, the von Neumann trace,
and the Maxwell-Boltzmann probability operator---
are well known
(see Refs.~[\onlinecite{Neumann27,Messiah61,Merzbacher70}]
in general, and
Refs.~[\onlinecite{Zurek82,Joos85,Zurek91,Zurek94,Zeh01,%
Zurek03,Zurek04,Schlosshauer05}]
for wave function collapse).
It is the derivation given in Paper I
that is of most interest.\cite{QSM1}

The second paper\cite{QSM2} derived the stochastic dissipative
Schr\"odinger equation for an open quantum system,
namely  a sub-system of a thermal reservoir,
the canonical equilibrium system.
Other modified versions of the Schr\"odinger equation exist
(e.g.\ Refs~[\onlinecite{Davies76,Kallianpur80,%
Gisin84,Belavkin89,Kummerer03,Bouten04,Pellegrini08}],
and also Refs~[\onlinecite{Breuer02,Weiss12}]).
The novelty of the stochastic dissipative Schr\"odinger equation
derived in Paper II\cite{QSM2}
was the first principles derivation based upon maximizing
the transition entropy.
Also, it was shown that the stochastic dissipative propagator
had to obey a quantum fluctuation dissipation theorem
that guaranteed the stationarity of the Maxwell-Boltzmann probability operator.

In a sense this, the third paper in the series,
is an intermezzo between the equilibrium and the non-equilibrium theory.
It addresses certain issues
that are required for the formulation of the non-equilibrium theory,
which  necessitates returning
to a first principles understanding
of the set theoretic basis for probability theory itself
in a way that is consistent with quantum theory.
This paper 
performs  the  analysis in an equilibrium  context.
Three new results emerge
beyond those already given in Papers I and II.\cite{QSM1,QSM2}
First,
in Paper II,\cite{QSM2} the stochastic dissipative time propagator
was taken to be unitary, for reasons of sufficiency and convenience.
Here it is shown that the unitary condition is a necessary
consequence of the reduction condition on the second entropy.
Second, the two-time operator product
of the forward and backward time propagators
is here shown to be equal to the conditional transition probability operator.
Third, a more exact understanding of the approximate nature
of the quadratic fluctuation form for the transition entropy operator
is developed in the present paper compared to Paper II.\cite{QSM2}
Beyond these equilibrium results that revisit Paper  II\cite{QSM1,QSM2}
---three steps backward as it were---
the set theoretic basis of non-equilibrium probability theory
is  summarized in \S \ref{Sec:neQSM}.
This is a small but necessary step toward the full non-equilibrium
theory that  is intended for the future.

\section{Nature of Probability in Quantum Systems} \label{Sec:Nat-Prob}
\setcounter{equation}{0}

\subsection{States}

In modeling a quantum system,
one has to specify the finest level of description that will be used,
which comes down to choosing a basis for the wave function.
Typically such a basis can be taken to be the set of eigenfunctions
of some operator, or set of commuting operators.
As far as the chosen level of description is concerned,
the basis and the operator are complete.
Of course in actual fact the real system may contain further degrees
of freedom or levels of description,
and one should have some reason for believing that these
are not directly relevant to the problem at hand
and can be neglected.

The finest level of description chosen will be called the microstates
of the system, and these correspond to the eigenstates of an operator.
Different operators yield different sets of eigenstates,
and different sets of eigenfunctions that serve as the basis
for the Hilbert space of the wave function.
Hence there is more than one way to represent the microstates
of a quantum system.
A microstate of one representation is a superposition of microstates
of another representation.
A complete set of commuting operators (a complete operator, for short)
yields non-degenerate eigenvalues, a distinct one for each microstate.

Macrostates may be defined by the principle quantum number
of the states of an incomplete set of commuting operators
(an incomplete operator, for short).
Such incomplete operators still have a set of corresponding microstates
or basis functions,
but in this case the microstates are degenerate,
with all microstates in each macrostate
having the same principle quantum number.

Let $\hat A = \hat A_a \hat A_b \ldots$
be an operator composed of commuting operators
that form a complete set.
\cite{Messiah61,Merzbacher70}
The case of most physical relevance is when the operators are Hermitian,
although this is not essential for the formal development of the theory.
The eigenvalue equation,
\begin{equation}
\hat A | \zeta_n^A \rangle = A_n | \zeta_n^A \rangle,
\end{equation}
defines the quantum states of $A$, $n=1,2,\ldots$.
The eigenfunctions form a complete orthonormal set,
$\langle \zeta_n^A |\zeta_m^A \rangle = \delta_{nm}$.
An arbitrary wave function may be expanded as
$|  \psi \rangle = \sum_n \psi_n^A | \zeta_n^A \rangle$,
with $\psi_n^A = \langle  \zeta_n^A | \psi \rangle$.

An important point is that for such a complete operator,
the quantum states are non-degenerate:
$A_n = A_m$ if and only if $n=m$.
The quantum states of such a complete operator
are the analogue of the classical microstates
in the set theoretic formulation of probability for statistical mechanics.
\cite{TDSM,NETDSM}

Let $\hat C$ be an Hermitian operator that is not complete.
The eigenvalue equation in this case may be written
\begin{equation}
\hat C | \zeta_{\alpha k}^C \rangle = C_\alpha | \zeta_{\alpha k}^C \rangle .
\end{equation}
The Greek index is the principle quantum number
and the Roman index labels the degeneracy.
The eigenfunctions again form a complete orthonormal set,
$\langle \zeta_{\alpha k}^C | \zeta_{\beta l}^C \rangle
= \delta_{\alpha\beta} \delta_{kl} $,
and
$|  \psi \rangle
= \sum_{\alpha,k} \psi_{\alpha k}^C | \zeta_{\alpha k}^C \rangle$,
with $\psi_{\alpha k}^C = \langle  \zeta_{\alpha k}^C | \psi \rangle$.
In this form  the principle quantum number uniquely specifies
the eigenvalue:
$C_\alpha = C_\beta$ if and only if $\alpha = \beta$.

For brevity, it is sometimes convenient
not to label the degeneracy explicitly
and to instead write,
\begin{equation}
\hat C | \xi_{\alpha}^C \rangle = C_\alpha | \xi_{\alpha}^C \rangle .
\end{equation}
In this form the convention invoked here is
that the principle quantum number still uniquely
specifies the eigenvalue,
$C_\alpha = C_\beta \Leftrightarrow \alpha = \beta$.
This means that the eigenfunction
$| \xi_{\alpha}^C \rangle$ belongs
to the sub-space spanned by  the degenerate eigenfunctions
of the  principle quantum number, $ | \zeta_{\alpha k}^C \rangle $.
The size of the sub-space
is the degeneracy of the macrostate,
which is the number of microstates it contains;
it is denoted $N_\alpha^C$.
The quantum states of such an incomplete operator labeled by the
principle quantum number
are the analogue of the classical macrostates
in the set theoretic formulation of probability for statistical mechanics.
\cite{TDSM,NETDSM}

\subsection{Weight}

The set theoretic formulation of probability in statistical mechanics
is based the existence of a weight for each microstate.\cite{TDSM,NETDSM}
The physical origin and value of such a weight is a question separate
to the formal development of the probability theory.
Here it can be mentioned
that  in Paper I\cite{QSM1} it was shown
that for an isolated quantum system,
the energy microstate weights are all equal,
and that without loss of generality
their value can be set to unity.
It was further shown that the weight operator
was diagonal in the basis of the complete energy operator.
From this one can conclude that the weight operator
of an isolated quantum system is the identity operator.

However, the most useful  application of quantum statistical mechanics
is to open quantum systems,
which is to say that one seeks to describe
in microscopic detail the behavior of a sub-system
that is interacting with a reservoir or environment
that  enters only through some macroscopic thermodynamic parameters.
In this case the microstates of the sub-system
do not have equal weight.
For this reason the  probability theory
for quantum statistical mechanics
is formulated in a general way
without assuming anything about the microstate weights
other than that they are real and positive.

Suppose that there is a weight operator $\hat w$
that is Hermitian, $\hat w^\dag = \hat w$.
In the equilibrium case, which is the focus of the present analysis,
the weight operator is also real, $\hat w^* = \hat w$
(see \S \ref{Sec:Parity}).
The matrix elements of the weight
in the representation of the complete operator $\hat A$
are
\begin{equation}
w_{mn}^A  \equiv
\frac{
\langle \zeta_m^A | \hat w |\zeta_n^A \rangle .
}{\sqrt{\langle \zeta_m^A | \zeta_m^A \rangle
\langle \zeta_n^A | \zeta_n^A \rangle } } .
\end{equation}
In this case the denominator is redundant because
the eigenfunctions are normalized.
The matrix is Hermitian, $w_{mn}^{A*} = w_{nm}^{A}$.

The diagonal elements are the weight of
the quantum states of the operator $\hat A$,
\begin{equation}
w_n^A
\equiv
w_{nn}^A
=
\frac{ \langle \zeta_n^A | \hat w |\zeta_n^A \rangle
}{ \langle \zeta_n^A | \zeta_n^A \rangle } .
\end{equation}
Since the weight matrix is Hermitian,
these diagonal elements are real.
They are also non-negative.
These diagonal elements of the representation of the weight
of a complete operator are the quantum analogue
of the weight of classical microstates.\cite{TDSM,NETDSM}
The off-diagonal elements have no classical analogue.

The total weight of the quantum system is
the sum of the weights of the quantum states,
\begin{equation} \label{Eq:W1}
W
= \sum_n w_n^A
= \sum_n
\frac{\langle \zeta_n^A | \hat w |\zeta_n^A \rangle
}{ \langle \zeta_n^A | \zeta_n^A \rangle }
= \mbox{Tr } \hat w.
\end{equation}
It will now be justified that this is indeed
independent of the operator $\hat A$.

Suppose that $\hat B$ is also a complete operator.
In this representation the elements of the weight  operator matrix are
\begin{eqnarray}
w_{kl}^B & = &
\langle \zeta_k^B | \hat w |\zeta_l^B \rangle
\nonumber \\ & = &
\sum_{mn}
\langle \zeta_k^B | \zeta_m^A \rangle
\, w_{mn}^A \,
\langle \zeta_n^A | \zeta_l^B \rangle .
\end{eqnarray}
The weight of the quantum states of $\hat B$ are
\begin{equation}
w_k^B \equiv   w_{kk}^B
=
\sum_{mn}
\langle \zeta_k^B | \zeta_m^A \rangle
\, w_{mn}^A \,
\langle \zeta_n^A | \zeta_k^B \rangle.
\end{equation}
The scalar product $\langle \zeta_n^A | \zeta_k^B \rangle$
represents the proportion of the quantum state
$\{n,A\}$ that appears in the quantum state  $\{k,B\}$,
and analogously for $\langle \zeta_k^B | \zeta_m^A \rangle$.
Hence this equation may be interpreted
as redistributing the weight from one representation to another.

The total weight in the representation $\hat B$ is
\begin{eqnarray}
W & = &
\sum_k w_k^B
\nonumber \\ & = &
\sum_k \sum_{mn}
\langle \zeta_k^B | \zeta_m^A \rangle
\, w_{mn}^A \,
\langle \zeta_n^A | \zeta_k^B \rangle
\nonumber \\ & = &
\sum_{mn}
\langle \zeta_n^A |  \zeta_m^A \rangle \, w_{mn}^A
\nonumber \\ & = &
\sum_{m}  w_{mm}^A
\nonumber \\ & = &
\mbox{Tr } \hat w .
\end{eqnarray}
This confirms that the total weight is independent of the representation.

As above,  $\hat C$ is an incomplete Hermitian operator
$ \hat C | \zeta_{\alpha k}^C \rangle
= C_\alpha | \zeta_{\alpha k}^C \rangle $.
One \emph{could} represent the weight operator
in the basis of $\hat C$,
with the matrix indexed by the principle quantum number
and the matrix elements defined as usual,
\begin{equation} \label{Eq:w_AlphaBeta}
w_{\alpha\beta}^C
\equiv
\frac{ \langle \xi_{\alpha}^C  | \hat w | \xi_{\beta}^C  \rangle
}{\sqrt{ \langle \xi_{\alpha}^C  | \xi_{\alpha}^C  \rangle
\langle \xi_{\beta}^C  |   \xi_{\beta}^C  \rangle } } .
\end{equation}
However this is of no direct use.
Alternatively, one could index the matrix by the microstates,
\begin{equation}
w_{\alpha k,\beta l}^C
\equiv
\frac{ \langle \zeta_{\alpha k}^C  | \hat w | \zeta_{\beta l}^C  \rangle
}{\sqrt{ \langle \zeta_{\alpha k}^C  | \zeta_{\alpha k}^C \rangle
\langle \zeta_{\beta l}^C  | \zeta_{\beta l}^C   \rangle } } .
\end{equation}
This is more reasonable,
and one may say that the weight of the microstates
of the operator $\hat C$ are
$w_{\alpha k}^C \equiv w_{\alpha k,\alpha k}^C$.

It is essential  that the weight of a principle quantum state
of such an incomplete operator is equal to the sum of the weight
of the degenerate quantum states that it contains,
\begin{eqnarray} \label{Eq:w-alpha-C}
w_{\alpha}^C
& \equiv &
\sum_{k\in\alpha}
w_{\alpha k,\alpha k}^C
\nonumber \\ & = &
\sum_{k\in\alpha}
\frac{ \langle \zeta_{\alpha k}^C  | \hat w | \zeta_{\alpha k}^C   \rangle
}{ \langle \zeta_{\alpha k}^C  | \zeta_{\alpha k}^C  \rangle  } .
\end{eqnarray}
Note that $w_{\alpha}^C \ne w_{\alpha\alpha}^C$,
where the right hand side is given by
the diagonal terms in Eq.~(\ref{Eq:w_AlphaBeta}).
In general the macrostate weight $w_{\alpha}^C$
cannot be written as a single expectation value
of the weight operator.

The total weight of the quantum system is the sum of the above,
\begin{equation}
W = \sum_\alpha w_{\alpha}^C
= \sum_{\alpha,k} w_{\alpha k,\alpha k}^C
= \mbox{Tr } \hat w .
\end{equation}
This is the total weight expressed as a sum over principle
quantum states (macrostates).
Again, $W \ne \sum_\alpha w_{\alpha\alpha}^C$,
using the diagonal terms in Eq.~(\ref{Eq:w_AlphaBeta}).
The second equality that expresses this
as a sum over microstates  confirms that the same total weight results
whether it is expressed in terms of microstates or macrostates.
For the total weight to be independent of whether
it invokes the weight associated with a complete or an incomplete
operator
(i.e.\ whether  it is summed over microstates or macrostates),
it is necessary that the definition
of principle quantum number weight, Eq.~(\ref{Eq:w-alpha-C}) be used.

\subsection{Probability}

One can define a probability operator in terms of the weight operator,
\begin{equation}
\hat \wp = \frac{1}{W} \hat w ,
\end{equation}
with normalization $\mbox{Tr } \hat \wp =1$.
This, the weight operator above,
and the entropy operator that follows are one-time operators.
On occasion the superscript `(1)' will be appended
in order to distinguish them more clearly from the two-time operators
that are introduced below for transitions.

The probability of a quantum state of a complete operator (microstate)
is its weight divided by the total weight,
\begin{equation}
\wp_n^A
= \frac{w_n^A}{W} .
\end{equation}
This is the same as the expectation value of the probability operator,
\begin{equation}
\wp_n^A
=
\frac{
\langle \zeta_n^A | \hat \wp  | \zeta_n^A \rangle
}{ \langle \zeta_n^A  | \zeta_n^A \rangle }
=
\frac{ \langle \zeta_n^A | \hat w  | \zeta_n^A \rangle
}{ \sum_n \langle \zeta_n^A  | \hat w  | \zeta_n^A \rangle } .
\end{equation}
By design, the microstate probability is normalized, $\sum_n \wp_n^A = 1$.
An alternative notation  is $\wp^A(n)$ or $\wp(n^A)$.

The probability of a principle quantum state of an incomplete operator
is
\begin{equation}
\wp_\alpha^C = \frac{w_\alpha^C }{W}.
\end{equation}
With the usual matrix representation,
$\wp_{\alpha k,\beta l}^C =
\langle \zeta_{\alpha k}^C | \hat \wp  | \zeta_{\beta l}^C \rangle$,
this can be written
\begin{eqnarray}
\wp_\alpha^C
& = &
\sum_{k\in\alpha}  \wp_{\alpha k,\alpha k}^C
\nonumber \\ & = &
\sum_{k\in\alpha}
\frac{
\langle \zeta_{\alpha k}^C | \hat \wp  | \zeta_{\alpha k}^C  \rangle
}{
\langle \zeta_{\alpha k}^C  | \zeta_{\alpha k}^C  \rangle}
\nonumber \\ & = &
\frac{
\sum_{k\in\alpha}
\langle \zeta_{\alpha k}^C | \hat w  | \zeta_{\alpha k}^C  \rangle
}{
 \sum_{\alpha,k}
\langle \zeta_{\alpha k}^C | \hat w  | \zeta_{\alpha k}^C  \rangle
} .
\end{eqnarray}

Note that the probability of
the principle quantum states of an incomplete operator
is not the expectation value of the probability operator
in that state,
\begin{equation}
\wp_\alpha^C
\ne
\frac{\langle \xi_{\alpha}^C | \hat \wp  | \xi_{\alpha}^C  \rangle
}{\langle \xi_{\alpha}^C  | \xi_{\alpha}^C  \rangle}
=
\frac{
 \sum_{k,l}\!\!\!\!\!\!^{(C,\alpha)} a_{\alpha k}^* a_{\alpha l}
\langle \zeta_{\alpha k}^C | \hat w  | \zeta_{\alpha l}^C  \rangle
}{
\sum_{\alpha} \sum_{k,l}\!\!\!\!\!\!^{(C,\alpha)} a_{\alpha k}^* a_{\alpha l}
\langle \zeta_{\alpha k}^C | \hat w  | \zeta_{\alpha l}^C  \rangle
} .
\end{equation}
This is erroneous because
the non-diagonal elements (superposition states) contribute to this.

Suppose that one has two incomplete operators,
$\hat C | \zeta_{\alpha k}^C \rangle = C_\alpha  | \zeta_{\alpha k}^C \rangle $
and
$\hat D | \zeta_{\beta l}^D \rangle = D_\beta  | \zeta_{\beta l}^D \rangle $,
and that these commute,
$\hat C \hat D = \hat D \hat C $.
They therefore  share eigenfunctions
and one can write
\begin{equation}
\hat C | \zeta_{\alpha\beta,k}^{CD} \rangle
=
C_\alpha | \zeta_{\alpha\beta,k}^{CD} \rangle
,\mbox{ and }
\hat D | \zeta_{\alpha\beta,k}^{CD}\rangle
=
D_\beta | \zeta_{\alpha\beta,k}^{CD} \rangle .
\end{equation}
Since the two operators commute,
the system can be in the state $\{ C, \alpha \}$
and in the state $\{ D, \beta \}$ simultaneously.
In this case the system is in  the common sub-space
$\{C,\alpha\} \cap \{D,\beta\}$.
The weight of such a simultaneous state is
\begin{eqnarray} \label{Eq:W^CD_alphabeta}
w^{CD}_{\alpha\beta}
& = &
\sum_{k \in \alpha\cap\beta }\!\!\!\!^{(CD)} \,
w_{\alpha\beta,k;\alpha\beta,k}^{CD}
\nonumber \\ & = &
\sum_{k \in \alpha\cap\beta }\!\!\!\!^{(CD)} \,
\frac{ \langle \zeta_{\alpha\beta,k}^{CD} |
\hat w
| \zeta_{\alpha\beta,k}^{CD} \rangle
}{\langle \zeta_{\alpha\beta,k}^{CD} | \zeta_{\alpha\beta,k}^{CD} \rangle
} .
\end{eqnarray}
An alternative notation for the left hand side is  $w(\alpha^{C}\beta^{D})$.

The principle quantum number labels a complete set.
Hence $\sum_\beta \alpha\cap\beta = \alpha$,
and $\sum_\alpha \alpha\cap\beta = \beta$.
This means that one has what might be called a conservation law for weight,
\begin{equation}
\sum_\beta\!^{(D)} w^{CD}_{\alpha\beta}
= w_\alpha^C
, \mbox{ and }
\sum_\alpha\!^{(C)}  w^{CD}_{\alpha\beta}
= w_\beta^D .
\end{equation}
Evidently then,
the sum of the joint weight over all pairs of
principle quantum numbers gives the total weight of the system,
\begin{equation}
\sum_{\alpha,\beta}\!^{(CD)} w^{CD}_{\alpha\beta}
= W .
\end{equation}

The unconditional probability that the quantum system
is simultaneously in the principle quantum states
$\{C,\alpha\}$ and $\{D,\beta\}$ is
\begin{equation}
\wp^{CD}(\alpha\beta)
= \frac{ w^{CD}_{\alpha\beta} }{ W },
\end{equation}
which can alternatively be written $\wp(\alpha^{C}\beta^{D})$.
If the system is in the  principle quantum state $\{D,\beta\}$,
then the conditional probability that it is also in the state $\{C,\alpha\}$
is
\begin{eqnarray} \label{Eq:wpCD-cond}
\wp^{CD}(\alpha|\beta)
& = &
\frac{1}{w_\beta^D}
\sum_{k \in \alpha\cap\beta }\!\!\!^{(CD)} \,
w_{\alpha\beta,k;\alpha\beta,k}^{CD}
\nonumber \\ & = &
\frac{w^{CD}_{\alpha\beta}}{W} \frac{W}{w_\beta^D}
\nonumber \\ & = &
\frac{\wp^{CD}(\alpha\beta)}{\wp_\beta^D} .
\end{eqnarray}
This is usually written in the form of Bayes' theorem,
\begin{equation}
\wp^{CD}(\alpha\beta)
= \wp^{CD}(\alpha|\beta) \, \wp_\beta^D ,
\end{equation}
or $\wp(\alpha^{C}\beta^{C})
= \wp(\alpha^{C}|\beta^{D}) \, \wp(\beta^{D}) $.

\subsection{Entropy\cite{NB1}}

The entropy of the total quantum system is
\begin{equation}
S = k_\mathrm{B} \ln W ,
\end{equation}
where $ k_\mathrm{B} = 1.38 \times 10^{-23}$\,J\,K$^{-1}$
is Boltzmann's constant.

Similarly,
the entropy of a microstate  of a complete operator  is
$S_n^A  =  k_\mathrm{B} \ln w_n^A $,
and that of a microstate of an incomplete operator is
$S_{\alpha k}^C  =  k_\mathrm{B} \ln w_{\alpha k}^C $.
In the same way, the entropy of a macrostate,
which is the principle quantum state of an incomplete operator, is
\begin{eqnarray}
S_\alpha^C
& = &
 k_\mathrm{B} \ln w_\alpha^C
\nonumber \\ & = &
k_\mathrm{B} \ln \sum_{k\in\alpha}
w_{\alpha k,\alpha k}^C .
\end{eqnarray}

One can relate the entropy operator to the weight operator by
\begin{equation}
\hat w = e^{ \hat S /k_\mathrm{B} }.
\end{equation}

As mentioned above, in general
the weight of a macrostate is not the expectation value
of the weight operator in that state.
Here can be added that neither is it the exponential of the expectation value
of the entropy operator in that state.
Further,
although the weight of a microstate \emph{is}  the expectation value
of the weight operator in that state,
in general it is \emph{not} the exponential of the expectation value
of the entropy operator in that state.
Because of this difference,
one needs to explicitly distinguish
the expectation value of the entropy operator,
\begin{equation}
S^{<>}_{mn} \equiv
\langle \zeta_{m}  | \hat S | \zeta_{n} \rangle ,
\end{equation}
from  the entropy as the logarithm of the weight,
\begin{equation}
S_{mn} \equiv
k_\mathrm{B} \ln w_{mn}
=
k_\mathrm{B} \ln
\langle \zeta_{m}  | \hat w | \zeta_{n} \rangle .
\end{equation}
When one speaks of the entropy of a state,
one has to be clear whether one means
$ S^{<>}_{nn} $ or  $ S_{nn} $.

The distinction between the two disappears
for the case of entropy microstates,
$  \hat S | \zeta_{\alpha k}^{S} \rangle
= S_\alpha | \zeta_{\alpha k}^{S} \rangle$.
In this case the expectation value of the exponential
is equal to the exponential of the expectation value,
\begin{eqnarray}
w_{\alpha k}^{S}
& = &
\langle \zeta_{\alpha k}^{S} | \hat w | \zeta_{\alpha k}^{S} \rangle
\nonumber \\ & = &
\langle \zeta_{\alpha k}^{S} | e^{ \hat S /k_\mathrm{B} }
| \zeta_{\alpha k}^{S} \rangle
\nonumber \\ & = &
\exp \left\{ \langle \zeta_{\alpha k}^{S} |
\hat S
| \zeta_{\alpha k}^{S} \rangle/k_\mathrm{B} \right\}
\nonumber \\ & = &
e^{S_\alpha /k_\mathrm{B}} .
\end{eqnarray}
In this case the two definitions of the entropy are
both equal  to the entropy eigenvalue,
$S_{\alpha k,\alpha k}^{S} = S^{S<>}_{\alpha k, \alpha k} = S_\alpha$.

The weight of the macrostate $\{\alpha,S\}$ is
\begin{eqnarray}
w_\alpha^S
& = &
\sum_{k\in\alpha}  w_{\alpha k,\alpha k}^S
\nonumber \\ & = &
N_\alpha^S
 e^{ S_{\alpha} /k_\mathrm{B} }
\nonumber \\ & = &
N_\alpha^S
\exp \frac{
\langle \xi_{\alpha}^{S} |\hat S | \xi_{\alpha}^{S}  \rangle
}{ k_\mathrm{B} \langle \xi_{\alpha}^{S} | \xi_{\alpha}^{S}  \rangle  }.
\end{eqnarray}
That is,
the weight of an entropy macrostate is
the number of degenerate states
times the exponential of the expectation value
of the entropy operator in that macrostate.

Obviously,
the entropy operator can be related to the probability operator,
\begin{equation}
\hat \wp = \frac{1}{W} e^{ \hat S /k_\mathrm{B} }.
\end{equation}
The probability of a state of a complete operator (microstate) is
\begin{eqnarray}
\wp_n^A & = &
\langle \zeta_{n}^{A} |\hat \wp | \zeta_{n}^{A} \rangle
 \\ \nonumber & = &
\frac{
\langle \zeta_{n}^{A} | e^{ \hat S /k_\mathrm{B} } | \zeta_{n}^{A} \rangle
}{
\sum_n
\langle \zeta_{n}^{A} | e^{ \hat S /k_\mathrm{B} } | \zeta_{n}^{A} \rangle
}
\nonumber \\ & \ne &
\frac{
\exp \left\{
\langle \zeta_{n}^{A} |\hat S | \zeta_{n}^{A}  \rangle
/ k_\mathrm{B}
\right\}
}{
\sum_n
\exp \left\{
\langle \zeta_{n}^{A} |\hat S | \zeta_{n}^{A}  \rangle
/ k_\mathrm{B}
\right\}
}
= 
\frac{
e^{ S_{nn}^{<>A} / k_\mathrm{B} }
}{
\sum_n e^{  S_{nn}^{<>A} / k_\mathrm{B} }
},
\nonumber
\end{eqnarray}
assuming a normalized basis,
$  \langle \zeta_{n}^{A} | \zeta_{n}^{A} \rangle = 1$.
The probability of a principle state of an incomplete operator (macrostate) is
\begin{eqnarray}
\wp_\alpha^C & = &
\sum_{k\in\alpha}
\langle \zeta_{\alpha k}^{C} |\hat \wp |  \zeta_{\alpha k}^{C} \rangle
 \\ \nonumber & = &
\frac{
\sum_{k\in\alpha}
 \langle \zeta_{\alpha k}^{C} |  e^{ \hat S /k_\mathrm{B} }
  |  \zeta_{\alpha k}^{C} \rangle
}{
\sum_{\alpha,k}
\langle \zeta_{\alpha k}^{C} |  e^{ \hat S /k_\mathrm{B} }
|  \zeta_{\alpha k}^{C} \rangle
}
\nonumber \\ & \ne &
\frac{
\sum_{k\in\alpha}
e^{
\langle \zeta_{\alpha k}^{C} |\hat S | \zeta_{\alpha k}^{C}  \rangle
/ k_\mathrm{B} }
}{
\sum_{\alpha,k}
e^{  \langle \zeta_{\alpha k}^{C} |\hat S | \zeta_{\alpha k}^{C} \rangle
/ k_\mathrm{B}  }
}
=
\frac{
\sum_{k\in\alpha}
e^{ S_{\alpha k,\alpha k}^{<>C} /k_\mathrm{B} }
}{
\sum_{\alpha,k} e^{ S_{\alpha k,\alpha k}^{<>C} /k_\mathrm{B} }
} , \nonumber
\end{eqnarray}
again assuming a normalized basis,
$ \langle  \zeta_{\alpha k}^{C} |  \zeta_{\alpha k}^{C} \rangle = 1$.
The error arises in both cases because
non-diagonal elements (superposition states)
preclude the interchange of expectation and exponentiation.

Since entropy eigenfunctions
are also probability  eigenfunctions,
$\hat \wp | \zeta_{\alpha k}^S \rangle
= \wp_{\alpha k}^S | \zeta_{\alpha k}^S \rangle
= W^{-1} e^{S_\alpha/k_\mathrm{B}} | \zeta_{\alpha k}^S \rangle $,
the probability operator matrix is diagonal
in the entropy representation,
$  \wp_{\alpha k, \beta l}^S
= \wp_{\alpha k}^S  \delta_{\alpha\beta} \delta_{kl}$.
(The degeneracy subscript $k$ is redundant on the probability eigenvalue.)
This means that
\begin{eqnarray}
\wp_{\alpha k}^S & = &
\langle \zeta_{\alpha k}^S |\hat \wp | \zeta_{\alpha k}^S \rangle
\nonumber \\  & = &
\frac{
\langle \zeta_{\alpha k}^S |
e^{ \hat S /k_\mathrm{B} } | \zeta_{\alpha k}^S \rangle
}{
\sum_{\alpha k}
\langle \zeta_{\alpha k}^S |
e^{ \hat S /k_\mathrm{B} } | \zeta_{\alpha k}^S \rangle
}
\nonumber \\ & = &
\frac{ e^{  S_\alpha  /k_\mathrm{B} }
}{
\sum_{\alpha k}  e^{  S_\alpha  /k_\mathrm{B} }
}
\nonumber \\ & = &
\frac{
e^{ S_{\alpha k, \alpha k}^{<>A} / k_\mathrm{B} }
}{
\sum_{\alpha k}  e^{  S_{\alpha k, \alpha k}^{<>A}/ k_\mathrm{B} }
}.
\end{eqnarray}
Further, the total entropy may be written
\begin{eqnarray}
S & = & k_\mathrm{B} \ln W
\nonumber \\ & = &
k_\mathrm{B} \sum_{\alpha,k} \wp_{\alpha k}^S  \ln W
\nonumber \\ & = &
- k_\mathrm{B} \sum_{\alpha,k} \wp_{\alpha k}^S
\left[ \ln \frac{e^{S_{\alpha k}^S/k_\mathrm{B}}}{W}
- S_{\alpha k}^S/k_\mathrm{B} \right]
\nonumber \\ & = &
- k_\mathrm{B} \sum_{\alpha,k} \wp_{\alpha k}^S
\left[ \ln \wp_{\alpha k}^S - S_{\alpha k}^S/k_\mathrm{B} \right]
\nonumber \\ & = &
- k_\mathrm{B} \sum_{\alpha,k}
\langle \zeta_{\alpha k}^S | \hat \wp
\left[ \ln \hat \wp  - \hat S /k_\mathrm{B} \right]
| \zeta_{\alpha k}^S \rangle
\nonumber \\ & = &
 \mbox{Tr } \hat \wp \, \hat S
 - k_\mathrm{B} \mbox{Tr } \hat \wp \ln \hat \wp .
\end{eqnarray}
The passage to the penultimate equality relied upon
the fact that the probability operator and the entropy operator
are diagonal in the entropy basis.
Although the entropy basis was used for the derivation,
the final expression is independent of any particular basis.
This exact formal expression for the total entropy of the quantum system
differs from what one finds in almost all other papers and text books.
The difference is the first term on the right hand side of the final equality,
whose physical interpretation is that of the internal entropy
of the macrostates.
This internal entropy of course contributes to
the total entropy of the system,
but unfortunately  conventional treatments
inadvertently neglect  it.

\subsection{Averages}

An arbitrary operator $\hat O$ has average value
\begin{eqnarray}
\left< \hat O \right>_\mathrm{stat}
& = &
\mbox{Tr } \hat \wp \, \hat O
\nonumber \\ & = &
\sum_{mn} \wp_{mn}^A \, O_{nm}^A
\nonumber \\ & = &
\sum_{\beta,l} \wp_{\beta l, \beta l}^O \, O_{\beta l, \beta l}^O
\nonumber \\ & = &
\sum_{\alpha,k} \wp_{\alpha k, \alpha k}^S \, O_{\alpha k, \alpha k}^S .
\end{eqnarray}
The final two equalities hold in representations
where one or other of the two operators is diagonal.
In such cases  $\wp_{\beta l, \beta l}^O \equiv  \wp_{\beta l}^O $
has the interpretation as the probability of the microstate
$\{O, \beta l\}$,
and $\wp_{\alpha k, \alpha k}^S \equiv \wp_{\alpha k}^S$
has the interpretation as the probability of the microstate
$\{S, \alpha k\}$.
In these two case the trace may be directly interpreted
as the weighted sum over microstates.
This form for the average results from the collapse of the wave function.
\cite{QSM1}

                \section{Transitions} \label{Sec:trans}
\setcounter{equation}{0}

\subsection{Transition Weight Operator}

Now consider transitions between the quantum states.
The transition $n_1^A \stackrel{\tau}{\rightarrow} n_2^B$
has microstate transition weight
$w_{n_2n_1}^{(2),BA}(\tau) \equiv w^{(2)}(n_2^{B},n_1^{A}|\tau)$.
In essence, this answers the question:
applying the operator $\hat A$ at time $t$,
and the operator $\hat B$ at time $t+\tau$,
what is the weight attached to  measuring $A_{n_1}$ and $B_{n_2}$?

One has to distinguish the present transition weight,
say between macrostates, $w^{(2)}(\beta^D,\alpha^C|\tau)$,
from the weight of the joint commuting macrostates,
$w _{\alpha\beta}^{CD} \equiv w(\alpha^C \beta^D)$,
as used in Eq.~(\ref{Eq:W^CD_alphabeta}) \emph{et seq.}
In the present case the operators do not necessarily commute,
and some care has to be exercised in taking the $\tau \rightarrow 0$ limit.

For the case of transitions in time $\tau$,
one defines the transition (or two-time)
weight operator $\hat w^{(2)}(\tau)$,
which can be represented as a double matrix with elements
\begin{equation}
w^{(2)}_{\stackrel{\scriptstyle B:m_2 n_2}{A:m_1 n_1}}(\tau)
=
\langle \zeta^B_{m_2} , \zeta^A_{m_1}  |
\hat w^{(2)}(\tau)
| \zeta^A_{n_1} , \zeta^B_{n_2}\rangle ,
\end{equation}
the basis functions being normalized.
The meaning and use
of such two-time operators will be clarified  below.

The total transition weight is
\begin{equation}
W^{(2)}(\tau)
= \mathrm{Tr}^{(2)}  \hat w^{(2)}(\tau)
= \sum_{n_2,n_1} w^{(2),BA}_{n_2n_1}(\tau) .
\end{equation}
The diagonal element that appears here,
$ w^{(2),BA}_{n_2n_1}(\tau) \equiv
w^{(2)}_{\stackrel{\scriptstyle B:n_2 n_2}{A:n_1 n_1}}(\tau)$,
is the weight attached to the transition
between states $n_1^A \stackrel{\tau}{\rightarrow} n_2^B$.
Obviously this result for the total transitions weight
is independent of the representation
in terms of the states of the operators
$\hat A$ and $\hat B$ in the second equality.
It will be shown in Eq.~(\ref{Eq:W2=W1}) below that
this total transition weight $W^{(2)}(\tau)$
is equal to the total weight $W$
(or more precisely $W^{(1)}$,
as given in Eq.~(\ref{Eq:W1})) for all $\tau$.

The second (or transition, or two-time) entropy operator
is simply defined as the logarithm of the transition weight operator,
\begin{equation}
\hat S^{(2)}(\tau) = k_\mathrm{B}  \ln \hat w^{(2)}(\tau) .
\end{equation}
The total transition entropy is of course
$ S^{(2)}(\tau) = k_\mathrm{B}  \ln W^{(2)}(\tau)$.
As in the case of the first entropy,
one has to distinguish between the expectation value
of the second entropy operator,
$S^{(2)<>}_{\ldots}(\tau)$,
and the logarithm of the expectation value of the transition weight operator,
$S^{(2)}_{\ldots}(\tau) = k_\mathrm{B}  \ln  w^{(2)}_{\ldots}(\tau) $.

The transition probability operator is
\begin{equation}
\hat \wp^{(2)}(\tau) =
\frac{\hat w^{(2)}(\tau)}{W^{(2)}(\tau)}
= \frac{e^{\hat S^{(2)}(\tau)/k_\mathrm{B} }}{W^{(2)}(\tau)} .
\end{equation}
This is obviously normalized,
$ \mathrm{Tr}^{(2)}  \hat \wp^{(2)}(\tau) = 1$.

The state transition probability is
\begin{equation}
\wp^{(2)}(n_2^B,n_1^A|\tau)
\equiv
\wp^{(2)}_{\stackrel{\scriptstyle B:n_2 n_2}{A:n_1 n_1}}(\tau)
= \frac{w^{(2),BA}_{n_2n_1}(\tau)}{W^{(2)}(\tau)} .
\end{equation}
This gives the unconditional probability of the transition
$n_1^A \stackrel{\tau}{\rightarrow} n_2^B$,
and it may also be written $\wp^{(2),BA}_{n_2n_1}(\tau)$.
Explicitly this is
\begin{equation}
\wp^{(2)}(n_2^B,n_1^A|\tau) =
\left< \zeta_{n_2}^B , \zeta_{n_1}^A \right|
\hat \wp^{(2)}(\tau)
\left| \zeta_{n_1}^A , \zeta_{n_2}^B \right> .
\end{equation}

The transition between two states is evidently analogous
to the joint probability of commuting macrostates,
and as in Eq.~(\ref{Eq:wpCD-cond}),
the conditional and the unconditional state transition probability
are related by Bayes' theorem,
\begin{equation} \label{Eq:wp2-cond}
\wp^{(2)}(n_2^B,n_1^A|\tau)
= \wp^{(2)}(n_2^B|n_1^A,\tau) \wp^{(1)}(n_1^A) ,
\end{equation}
where $\wp^{(1)}(n_1^A) \equiv \wp^{A}_{n_1}$.
The conditional  state transition probability
$\wp^{(2)}(n_2^B|n_1^A,\tau)$  is the probability of the system being
in the state $n_2^B$ at time $t_2 = t_1+\tau$
given that it is in the state ${n_1^A}$ at time $t_1$.
The time interval $\tau$ may be positive or negative;
if positive, then this is the probability
of the transition from the current state,
and if negative, then this is the probability of  the transition
that led to the current state.
Since
$\wp^{(2)}(n_2^B,n_1^A|\tau) = \wp^{(2)}(n_1^A,n_2^B|{-\tau})$
(time homogeneity combined with statistical symmetry),
one must also have
\begin{equation}
\wp^{(2)}(n_2^B,n_1^A|\tau) =
\wp^{(2)}(n_1^A|n_2^B,-\tau) \wp^{(1)}(n_2^B) .
\end{equation}
The conditional transition probability is conditioned on  pure quantum states,
which is to say that it is  not defined here
for off-diagonal conditioning states.

The evolution of the present open equilibrium quantum system
can also be characterized by a stochastic dissipative
time propagator, $\hat{\cal U}(\tau)$,
\begin{equation}
| \psi(t+\tau) \rangle
= \hat{\cal U}(\tau) | \psi(t) \rangle .
\end{equation}
This time propagator must obey certain statistical rules and symmetries
that will be derived below.
The conditional state transition probability can be expressed in terms
of the propagator, namely
\begin{eqnarray} \label{Eq:wp2-UUb}
\lefteqn{
\wp^{(2)}(n_2^B|n_1^A,\tau)
}
\nonumber \\  & =&
\left<
{\cal U}_{n_1,n_2}^{AB}(\tau)^\dag {\cal U}_{n_2,n_1}^{BA}(\tau)
\right>_\mathrm{stoch}
 \nonumber \\  & =&
\left<
\langle \zeta_{n_1}^A  |  \hat{\cal U}(\tau)^\dag | \zeta_{n_2}^B  \rangle
\,
\langle \zeta_{n_2}^B |  \hat{\cal U}(\tau) | \zeta_{n_1}^A \rangle
\right>_\mathrm{stoch}.
\end{eqnarray}
The angular brackets signify an average over the stochastic propagator.
This propagator form for the conditional state transition probability
is motivated by the analysis of the time correlation function
in the section that follows.

The  conditional transition probability operator
is a two-time operator,
and writing it as the composition of the two one-time time propagators
is a notational challenge that is perhaps best met by
\begin{equation}
\hat \wp^{(2),\mathrm{cond}}(\tau)
=
\left< \left\{
\hat {\cal U}(\tau)^\dag ,  \hat {\cal U}(\tau)
\right\} \right>_\mathrm{stoch}  .
\end{equation}
The meaning of the notation can  be best gauged from the representation,
\begin{equation}
\wp^{(2),\mathrm{cond}}_{\stackrel{\scriptstyle m_2 n_2}{m_1 n_1}}(\tau)
=
\left< {\cal U}_{m_1n_2}(\tau)^\dag
{\cal U}_{m_2n_1}(\tau) \right>_\mathrm{stoch} .
\end{equation}

The unconditional transition probability operator
is the composition
of the conditional transition probability operator
and the singlet probability operator,
which can be arranged in four ways,
\begin{eqnarray} \label{Eq:hatUUwp1}
\hat \wp^{(2)}(\tau)
& = &
\left< \left\{
\hat {\cal U}(\tau)^\dag ,  \hat {\cal U}(\tau) \hat \wp^{(1)}
\right\} \right>_\mathrm{stoch}
\nonumber \\ & = &
\left< \left\{
\hat \wp^{(1)} \hat {\cal U}(\tau)^\dag ,  \hat {\cal U}(\tau)
\right\} \right>_\mathrm{stoch}
\nonumber \\ & = &
\left< \left\{
\hat {\cal U}(\tau)^\dag , \hat \wp^{(1)} \hat {\cal U}(\tau)
\right\} \right>_\mathrm{stoch}
\nonumber \\ & = &
\left< \left\{
\hat {\cal U}(\tau)^\dag \hat \wp^{(1)} ,  \hat {\cal U}(\tau)
\right\} \right>_\mathrm{stoch} ,
\end{eqnarray}
recalling that $\hat \wp^{(1)\dag} = \hat \wp^{(1)}$.
In these the singlet probability is combined with a time propagator
using the ordinary one-time composition of operators.
Hence one has four representations of the unconditional transition probability,
\begin{eqnarray}  \label{Eq:hatUUwp1-repn}
\wp^{(2)}_{\stackrel{\scriptstyle m_2 n_2}{m_1 n_1}}(\tau)
& = &
\sum_{\stackrel{\scriptstyle m_2 ,n_2}{m_1, n_1, l_1}}
\left< {\cal U}_{m_1n_2}(\tau)^\dag
{\cal U}_{m_2l_1}(\tau) \wp^{(1)}_{l_1 n_1} \right>_\mathrm{stoch} ,
\nonumber \\ & = &
\sum_{\stackrel{\scriptstyle m_2, n_2}{m_1, n_1, l_1}}
\left< \wp^{(1)}_{m_1 l_1} {\cal U}_{l_1n_2}(\tau)^\dag
{\cal U}_{m_2n_1}(\tau) \right>_\mathrm{stoch} ,
\nonumber \\ & = &
\sum_{\stackrel{\scriptstyle m_2, n_2, l_2}{m_1, n_1 }}
\left< {\cal U}_{m_1n_2}(\tau)^\dag
\wp^{(1)}_{m_2 l_2} {\cal U}_{l_2n_1}(\tau)  \right>_\mathrm{stoch} ,
\nonumber \\ & = &
\sum_{\stackrel{\scriptstyle m_2, n_2, l_2}{m_1, n_1 }}
\left<{\cal U}_{m_1 l_2}(\tau)^\dag \wp^{(1)}_{l_2 n_2}
 {\cal U}_{m_2n_1}(\tau)  \right>_\mathrm{stoch} ,
 \nonumber \\
\end{eqnarray}
respectively.
As usual,
${\cal U}_{m_1n_2}(\tau)^\dag = {\cal U}_{n_2m_1}(\tau)^*$.

If the system is in the wave state $\psi_1$ at $t_1$
and $\psi_2$ at time $t_2=t_1+\tau$,
then the expectation value of the transition probability is
\begin{eqnarray}
\lefteqn{
\wp^{(2)}(\psi_2,\psi_1|\tau)
} \nonumber \\
& = &
\frac{\langle \psi_2,\psi_1 | \hat \wp^{(2)} |  \psi_1,\psi_2 \rangle
}{
\langle \psi_2| \psi_2 \rangle \, \langle \psi_1| \psi_1 \rangle}
\nonumber \\ & = &
\frac{1}{N(\psi_2) N(\psi_1)}
\sum_{\stackrel{\scriptstyle m_2 n_2}{m_1 n_1}}
\psi_{2,m_2}^* \psi_{2,n_2} \psi_{1,m_1}^* \psi_{1,n_1}
\wp^{(2)}_{\stackrel{\scriptstyle m_2 n_2}{m_1 n_1}}(\tau)
\nonumber \\ & = &
\frac{1}{N(\psi_1)N(\psi_2)}
\sum_{\stackrel{\scriptstyle m_2,n_2}{m_1,n_1,l_1}}
\psi_{2,m_2}^* \psi_{2,n_2}  \psi_{1,m_1}^* \psi_{1,n_1}
\nonumber \\ && \mbox{ } \times
\left< {\cal U}_{m_1n_2}(\tau)^\dag
{\cal U}_{m_2l_1}(\tau) \right>_\mathrm{stoch}
\wp^{(1)}_{l_1n_1}.
\end{eqnarray}
There are three other versions of the final equality
that can be derived from the preceding equation.
As will be shown next, in practice one deals with fully collapsed states,
$m_2 = n_2$ and $l_1= m_1 = n_1$.

\subsection{Time Correlation Function}

In order to understand the meaning of a two-time operator,
one can look at the correlation of two operators,
$\hat B$ applied at time $t_2 = t_1+\tau$ and $\hat A$ applied at time $t_1$.
The time correlation function is
\begin{eqnarray} \label{Eq:CBA-gen}
C_{BA}(\tau)
& \equiv &
\mathrm{Tr}^{(2)}
\left[ \hat \wp^{(2)}(\tau) \{ \hat B(t_1+\tau), \, \hat A(t_1) \} \right]
\nonumber \\ & = &
\sum_{\stackrel{\scriptstyle m_2,n_2}{m_1,n_1}}
 \wp^{(2)}_{\stackrel{\scriptstyle m_2 n_2}{m_1 n_1}}(\tau)
 B_{n_2,m_2}  A_{n_1,m_1} .
\end{eqnarray}
If one uses the basis $\{ \zeta_n^A \}$ at $t_1$
and $\{ \zeta_n^B \}$ at $t_2 = t_1+\tau$,
then the operator matrices are diagonal and this becomes
\begin{eqnarray}
C_{BA}(\tau)
& = &
\sum_{n_2,n_1}
\wp^{(2)}_{\stackrel{\scriptstyle B:n_2 n_2}{A:n_1 n_1}}(\tau)
B_{n_2 n_2}^B  A_{n_1 n_1}^A
\nonumber \\ & \equiv &
\sum_{n_2,n_1}
\wp^{(2),BA}_{n_2 n_1}(\tau) B_{n_2}  A_{n_1}  .
\end{eqnarray}
Since
$\wp^{(2),BA}_{n_2n_1}(\tau)$
is the unconditional probability of the transition
between states of the two operators,
$n_1^A \stackrel{\tau}{\rightarrow } n_2^B$,
this second form emphasizes the collapse of the wave function
into the pure quantum states of the two operators.

Using the propagator expression in the preceding section
for the conditional transition probability, Eq.~(\ref{Eq:hatUUwp1-repn}),
and combining it with the notion of collapse into the pure quantum states,
the time correlation function
can also be written in terms of the propagator,
\begin{eqnarray}
\lefteqn{
C_{BA}(\tau)
} \nonumber \\
& = &
\!\!\!
\sum_{\stackrel{\scriptstyle m_2,n_2}{m_1,n_1,l_1}} \!\!\!
\left< {\cal U}_{m_1n_2}(\tau)^\dag
{\cal U}_{m_2l_1}(\tau) \right>_\mathrm{stoch}
\wp^{(1)}_{l_1 n_1}
B_{n_2m_2} A_{n_1m_1}
\nonumber \\ & \stackrel{\mbox{coll.}}{\Rightarrow} &
\sum_{n_1,n_2}
\left< {\cal U}_{n_1n_2}^{AB}(\tau)^\dag B_{n_2n_2}^B
{\cal U}_{n_2n_1}^{BA}(\tau)  \wp_{n_1n_1}^A A_{n_1n_1}^A
\right>_\mathrm{stoch}
\nonumber \\ & =&
\sum_{n_1,n_2}
\left< {\cal U}_{n_1n_2}^{AB}(\tau)^\dag B_{n_2}
{\cal U}_{n_2n_1}^{BA}(\tau)  \wp_{n_1}^A A_{n_1}
\right>_\mathrm{stoch}.
\end{eqnarray}
The operator $\hat A$ has collapsed the probability operator
from superposition states into pure states,
$\wp_{l_1n_1}^A \Rightarrow \wp_{n_1n_1}^A \equiv  \wp_{n_1}^A$.
This has a straightforward physical interpretation
for the meaning of the time correlation function,
which adds credibility to Eq.~(\ref{Eq:wp2-UUb}).
In words, $ \wp_{n_1}^A$ is the probability of measuring $A_{n_1}$,
and, given the state $n_1^A$,
$ {\cal U}_{n_1n_2}^{AB}(\tau)^\dag {\cal U}_{n_2n_1}^{BA}(\tau) $
is the probability of making the transition to the state $n_2^B$
and thence measuring $B_{n_2}$.
The expression for the conditional transition probability,
Eq.~(\ref{Eq:hatUUwp1}),
can only be applied to two-time averages in the context of such a collapse.

A second version of the time correlation function
results upon replacing $\wp_{n_1}^A$ by $\wp_{n_2}^B$
in the final equality.

If one sets $\hat B \equiv \hat{\mathrm I}$,
then the time correlation function is just  the average value
of the operator  $\hat A$,
\begin{equation}
C_{{\mathrm I}A}(\tau) =
\left< \hat A \right>_\mathrm{stat}
= \mbox{Tr } \hat \wp^{(1)} \, \hat A
= \sum_{mn} \wp^{(1)}_{mn} A_{nm} .
\end{equation}
Prior to the collapse the time correlation function is explicitly
\begin{eqnarray}
\lefteqn{
C_{{\mathrm I}A}(\tau)
} \nonumber \\
& = &
\sum_{n_2} \sum_{m_1,n_1,l_1}
\left< {\cal U}_{m_1n_2}(\tau)^\dag
{\cal U}_{n_2l_1}(\tau) \right>_\mathrm{stoch}
\wp^{(1)}_{l_1 n_1}
 A_{n_1m_1}
\nonumber \\ & = &
\sum_{m_1,n_1}
\left\{ \left< \hat {\cal U}(\tau)^\dag
\hat {\cal U}(\tau)\right>_\mathrm{stoch} \hat \wp^{(1)}
\right\}_{m_1n_1}
A_{n_1m_1}
\end{eqnarray}
Equating these two yields
\begin{equation}
\hat \wp^{(1)}
=
\left< \hat {\cal U}(\tau)^\dag
\hat {\cal U}(\tau) \right>_\mathrm{stoch} \hat \wp^{(1)},
\end{equation}
or
\begin{equation}  \label{Eq:UU=I}
\left< \hat {\cal U}(\tau)^\dag
\hat {\cal U}(\tau) \right>_\mathrm{stoch}
=
\hat{ \mathrm I} .
\end{equation}
(The probability operator can be taken in or out of the stochastic average
as desired.)
On the left hand side is the ordinary composition of one-time operators.
This says that the time propagator must on average be unity.

It is an arguable point whether one should equate the two expressions
for the time correlation function before or after the collapse.
If it is done after the collapse,
then one obtains the weaker result that says that the diagonal elements
are unity, but is silent about the non-diagonal elements.
The point will be rendered moot in the following section
when the general result and its variants will be derived
in a more rigorous fashion.

Using the relationship between the conditional probability
and the propagators,
this last result shows that
the conditional state probability must sum to unity,
\begin{equation} \label{Eq:sum-wp2-cond}
\sum_{n_2} \wp^{(2)}(n_2|n_1,\tau) = 1  .
\end{equation}
As for the unconditional probability,
inserting $\hat B \equiv \hat{\mathrm I}$
into Eq.~(\ref{Eq:CBA-gen})
and comparing it to $\langle \hat A \rangle_\mathrm{stat}$
one obtains
\begin{equation}
\sum_{n_2}  \wp^{(2)}_{\stackrel{\scriptstyle n_2 n_2}{m_1 n_1}}(\tau)
= \wp^{(1)}_{m_1n_1}  .
\end{equation}
This and the preceding two results will now be obtained
by an alternative  method.

\subsection{Reduction Condition}

Since the microstates are complete,
one must have the reduction conditions,
\begin{equation}
\sum_{n_2^B} w_{n_2^B n_1^A}^{(2)}(\tau) = w_{n_1^A}
,\mbox{ and }
\sum_{n_1^A} w_{n_2^B n_1^A}^{(2)}(\tau) = w_{n_2^B} .
\end{equation}
Recall the variant notations:
$w^{(2)}_{n_2^B n_1^A}(\tau) \equiv w_{n_2 n_1}^{(2),BA}(\tau)
\equiv w^{(2)}(n_2^B,n_1^A|\tau)$,
and $w_{n_1^A} \equiv w_{n_1}^A \equiv w_{n_1}^{(1),A} \equiv w^{(1)}(n_1^A)$.

It follows that the total transition weight
is equal to the total weight, $W^{(1)} \equiv W$,
\begin{equation} \label{Eq:W2=W1}
W^{(2)}(\tau)
= \sum_{n_2^B,n_1^A} w_{n_2^B n_1^A}^{(2)}(\tau)
= \sum_{n_1^A} w_{n_1^A}
= W.
\end{equation}
These are a type of conservation law for weight,
and the results only hold for the present equilibrium system.
The reduction also occurs in macrostates,
\begin{equation}
\sum_{\alpha_2^D} w_{\alpha_2^D \beta_1^C}^{(2)}(\tau)
=
w_{\beta_1^C },
\mbox{ and }
\sum_{\beta_1^C } w_{\alpha_2^D \beta_1^C}^{(2)}(\tau)
=
w_{\alpha_2^D}.
\end{equation}

The reduction condition on microstate weights
leads to a reduction condition on the weight operators,
\begin{equation} 
\mbox{Tr}^{(1)} \hat w^{(2)}(\tau)
=
\hat w^{(1)} .
\end{equation}
The notation signifies the partial trace
over one of the two times of the operator;
this could be made a little clearer by signifying explicitly
which one of the two times were traced over.
Explicitly, applying the trace to the second time this is
\begin{eqnarray}
\sum_{n_2}
w^{(2)}_{\stackrel{\scriptstyle B:n_2 n_2}{A:m_1 n_1}}(\tau)
& = &
\sum_{n_2} \langle \zeta^B_{n_2} , \zeta^A_{m_1}  |
\hat w^{(2)}(\tau)
| \zeta^A_{n_1} , \zeta^B_{n_2}\rangle ,
\nonumber \\ & = &
\langle  \zeta^A_{m_1}  |
\hat w^{(1)}(\tau)
| \zeta^A_{n_1} \rangle
\nonumber \\ & = &
w_{m_1,n_1}^{(1),A} .
\end{eqnarray}
(The superscript `A' rather than the more precise but not particularly
useful `AA' is used for the representation of one-time operators here and below.)
Alternatively, applying the trace to the first time it is
\begin{eqnarray}
\sum_{n_1}
w^{(2)}_{\stackrel{\scriptstyle B:m_2 n_2}{A:n_1 n_1}}(\tau)
& = &
\sum_{n_1} \langle \zeta^B_{m_2} , \zeta^A_{n_1}  |
\hat w^{(2)}(\tau)
| \zeta^A_{n_1} , \zeta^B_{n_2}\rangle ,
\nonumber \\ & = &
\langle  \zeta^B_{m_2}  |
\hat w^{(1)}(\tau)
| \zeta^B_{n_2} \rangle
\nonumber \\ & = &
w_{m_2,n_2}^{(1),B} .
\end{eqnarray}
These generalize the reduction condition
for the transition weight to off-diagonal elements.

Because the total transition weight is also the total weight,
$W^{(2)}(\tau) = W$,
the reduction condition for the weight operators
carry over directly to the probability operators,
\begin{equation} \label{Eq:wp1=red(wp2)}
\mbox{Tr}^{(1)} \hat \wp^{(2)}(\tau)
=
\hat \wp^{(1)} .
\end{equation}
The general (off-diagonal) results from this are
\begin{equation}
\sum_{n_2}
\wp^{(2)}_{\stackrel{\scriptstyle B:n_2 n_2}{A:m_1 n_1}}(\tau)
= \wp_{m_1,n_1}^{(1),A} ,
\end{equation}
and
\begin{equation}
\sum_{n_1}
\wp^{(2)}_{\stackrel{\scriptstyle B:m_2 n_2}{A:n_1 n_1}}(\tau)
= \wp_{m_2,n_2}^{(1),B}.
\end{equation}
In terms of the transitions between the operator microstates these become,
\begin{equation}
\sum_{n_2^B} \wp_{n_2^B n_1^A}^{(2)}(\tau) = \wp_{n_1^A}^{(1)}
,\mbox{ and }
\sum_{n_1^A} \wp_{n_2^B n_1^A}^{(2)}(\tau) = \wp_{n_2^B}^{(1)} ,
\end{equation}
and similarly for macrostates.

For the conditional state transition probability,
the reduction becomes
\begin{equation}
\sum_{n_2^B} \wp(n_2^B|n_1^A,\tau) = 1
,\mbox{ and }
\sum_{n_1^A} \wp({n_1^A}|n_2^B,-\tau) = 1 .
\end{equation}
This confirms Eq.~(\ref{Eq:sum-wp2-cond}).
Explicitly, in terms of the propagator this is
\begin{eqnarray}
1 & = &
\sum_{n_2^B}
\left<
\langle \zeta_{n_1}^A | \hat{\cal U}(\tau)^\dag | \zeta_{n_2}^B \rangle
\langle \zeta_{n_2}^B  | \hat{\cal U}(\tau) | \zeta_{n_1}^A \rangle
\right>_\mathrm{stoch}
\nonumber \\ & = &
\sum_{n_2^B}
\left< {\cal U}_{n_1n_2}^{AB}(\tau)^\dag
\hat{\cal U}_{n_2n_1}^{BA}(\tau)  \right>_\mathrm{stoch}
\nonumber \\ & = &
\left<
\langle \zeta_{n_1}^A | \hat{\cal U}(\tau)^\dag
\hat{\cal U}(\tau) | \zeta_{n_1}^A \rangle
\right>_\mathrm{stoch} .
\end{eqnarray}
These are just the diagonal elements of Eq.~(\ref{Eq:UU=I}).

Inserting the propagator expressions for the  transition probability operator,
Eq.~(\ref{Eq:hatUUwp1-repn}),
into the reduction condition, Eq.~(\ref{Eq:wp1=red(wp2)}),
and taking the trace over $t_2$ yields
\begin{eqnarray}
\hat \wp^{(1)}
& = &
\left<
\hat {\cal U}(\tau)^\dag  \hat {\cal U}(\tau) \hat \wp^{(1)}
\right>_\mathrm{stoch}
\nonumber \\ & = &
\left<
\hat \wp^{(1)} \hat {\cal U}(\tau)^\dag \hat {\cal U}(\tau)
\right>_\mathrm{stoch}
\nonumber \\ & = &
\left<
\hat {\cal U}(\tau)^\dag \hat \wp^{(1)} \hat {\cal U}(\tau)
\right>_\mathrm{stoch}
\nonumber \\ & = &
\left<
\hat {\cal U}(\tau)^\dag \hat \wp^{(1)}   \hat {\cal U}(\tau)
\right>_\mathrm{stoch} .
\end{eqnarray}
In this case the trace over the second time has become
the ordinary one-time composition of operators.
Taking the trace over the first time yields
\begin{eqnarray}
\hat \wp^{(1)}
& = &
\left<  \hat {\cal U}(\tau) \hat \wp^{(1)} \hat {\cal U}(\tau)^\dag
\right>_\mathrm{stoch}
\nonumber \\ & = &
\left<
\hat {\cal U}(\tau) \hat \wp^{(1)} \hat {\cal U}(\tau)^\dag
\right>_\mathrm{stoch}
\nonumber \\ & = &
\left<
\hat \wp^{(1)} \hat {\cal U}(\tau) \hat {\cal U}(\tau)^\dag
\right>_\mathrm{stoch}
\nonumber \\ & = &
\left<
\hat {\cal U}(\tau) \hat {\cal U}(\tau)^\dag \hat \wp^{(1)}
\right>_\mathrm{stoch} .
\end{eqnarray}
In this case the trace over $t_1$ has rearranged the order of
the operators in order to express the result
as the ordinary one-time composition of the operators.

From the first two equations of the first set,
and from the last two equations of the second set,
one sees that
\begin{equation} \label{Eq:UU=I:proof}
\left< \hat {\cal U}(\tau)^\dag  \, \hat {\cal U}(\tau) \right>_\mathrm{stoch}
=
\left< \hat {\cal U}(\tau) \, \hat {\cal U}(\tau)^\dag  \right>_\mathrm{stoch}
=
\hat{\mathrm I} .
\end{equation}
This says that on average
the stochastic dissipative propagator is unitary,
just as in the  adiabatic case.
This confirms Eq.~(\ref{Eq:UU=I}).

Because the identity operator  is transpose symmetric
(the same result follows by invoking its reality),
it follows from these that
\begin{eqnarray}
\langle \hat {\cal U}(\tau)^\dag \, \hat {\cal U}(\tau) \rangle_\mathrm{stoch}
& = &
\langle \hat{\cal U} (\tau)^\mathrm{T}
\hat {\cal U}(\tau)^* \rangle_\mathrm{stoch}
\nonumber \\ & = &
\langle \hat{\cal U} (\tau)^*
\hat {\cal U}(\tau)^\mathrm{T} \rangle_\mathrm{stoch}  .
\end{eqnarray}
For the second equality to equal the left hand side,
it is sufficient that the propagator be transpose symmetric.
It is certainly a convenience for the propagator to be transpose symmetric,
but the author has not been able to develop a mathematical proof
or a physical argument that this is in fact a necessary condition.

The remaining equations of the reduction condition
correspond to the stationarity of the probability operator
under the action of the propagator,
\begin{eqnarray} \label{Eq:wp-stationary}
\hat \wp^{(1)}
& = &
\left<
\hat {\cal U}(\tau)^\dag \hat \wp^{(1)} \hat {\cal U}(\tau)
\right>_\mathrm{stoch}
\nonumber \\ & = &
\left<  \hat {\cal U}(\tau) \hat \wp^{(1)} \hat {\cal U}(\tau)^\dag
\right>_\mathrm{stoch} .
\end{eqnarray}
Since the singlet probability operator is transpose symmetric,
(the same result follows by invoking its reality),
one also has
\begin{eqnarray}
\hat \wp^{(1)}
& = &
\left<
\hat {\cal U}(\tau)^\mathrm{T} \hat \wp^{(1)} \hat {\cal U}(\tau)^*
\right>_\mathrm{stoch}
\nonumber \\ & = &
\left<  \hat {\cal U}(\tau)^* \hat \wp^{(1)} \hat {\cal U}(\tau)^\mathrm{T}
\right>_\mathrm{stoch} .
\end{eqnarray}
It is sufficient, but not necessary,
for the propagator and the probability operator to commute
in order for these stationarity results
to follow directly from the unitary condition.

The reduction condition for the second entropy
may be rewritten as
\begin{eqnarray} \label{Eq:S1n=S2nm-red}
S^A_{n_1} & = & k_\mathrm{B} \ln w^A_{n_1}
\nonumber \\ & = &
k_\mathrm{B} \ln \sum_{n_2} w^{BA}_{n_2n_1}(\tau)
\nonumber \\ & \approx &
k_\mathrm{B} \ln  w^{BA}_{\overline n_2n_1}(\tau)
\nonumber \\ & = &
S^{(2),BA}_{\overline n_2n_1}(\tau) .
\end{eqnarray}
Here $ \overline n_2  \equiv  \overline n_2^B(\tau|n_1^A)$
denotes the most likely destination state of the operator $\hat B$
for the transition from the initial state $ n_1^A$
in time $\tau$.
This most likely state is the one that maximizes the second entropy,
\begin{equation}
\left. \frac{\partial S^{(2),BA}_{n_2,n_1}(\tau)}{\partial n_2}
\right|_{n_2= \overline n_2}
= 0,
\;\; \overline n_2  \equiv  \overline n_2^B(\tau|n_1^A).
\end{equation}
This is of course the same as maximizing the transition weight,
or as maximizing the transition probability.
The third equality above follows because the weight distribution
may be expected to be sharply peaked,
and consequently the logarithm of the sum
is equal to the logarithm of the largest term in the sum.

\subsection{Parity and Reversibility}

\subsubsection{Parity} \label{Sec:Parity}

In general the complex conjugate corresponds to velocity reversal.
Every microstate $n$ has a conjugate state  $n^\dag$
that is the same state  but with the velocities reversed.

A real operator, $\hat A^* =\hat A$,
is said to have even parity
because its eigenvalues are insensitive to the sign of the velocity.
This means that the eigenvalues of the conjugate operator,
$\hat A^* |\zeta_n^A \rangle = A_{n^\dag}  |\zeta_n^A \rangle$,
and that of the original,
$\hat A |\zeta_n^A \rangle = A_{n}  |\zeta_n^A \rangle$,
are equal $ A_{n^\dag} = A_n$.
In general microstates are defined such that their eigenvalues are
distinct (i.e.\ non-degenerate).
In the case of an even parity operator,
this means that the conjugate and the original state are the same state,
$\{ A,n^\dag \} \equiv \{ A,n \}$.

An odd parity operator is imaginary,
$\hat B^* = - \hat B$, and
$\hat B^* |\zeta_n^B \rangle
\equiv B_{n^\dag}  |\zeta_n^B \rangle
= -B_n  |\zeta_n^B \rangle$.
In this case the eigenvalues  are different,
$ B_{n^\dag} = -B_n$,
and the conjugate state is different to the original state,
$\{ B,n^\dag \} \ne \{ B,n \}$.
Taking the complex conjugate of the eigenvalue equation
shows that
$\hat B | \zeta_n^{B*} \rangle = -B_n | \zeta_n^{B*} \rangle$,
and hence $| \zeta_n^{B*} \rangle = | \zeta_{n^\dag}^{B} \rangle$,
and that this is a distinct eigenfunction of the operator $\hat B$.

Whether an operator is real or imaginary or complex,
one always has the conjugate state $n^\dag$.
It is only in the real case that this is
the same as the original state.
In general (i.e.\ for an even, odd, or complex operator),
$\langle \zeta_{n}| \zeta_{n^\dag} \rangle
= \delta_{n,n^\dag}$.

When one writes $\sum_n$,
both $n$ and $n^\dag$ are included in the sum.
For an even parity operator these are the same state
and it is included once only,
whereas for an odd parity operator they are distinct.
Since $n$ is a dummy summation index one can freely write
$\sum_n f_n = \sum_{n^\dag} f_n = \sum_n f_{n^\dag}$.
Obviously $(n^\dag)^\dag = n$.

The set of eigenfunctions of a complete operator
forms a complete basis,
which means that there is the same number (possibly infinite)
of microstates for all operators.
For example,
the sum over the microstates of an even parity operator might be written,
$\sum_n\!\!\!^{(A)} = \sum_{n=1}^{2N}$,
whereas the sum over the microstates
of an odd parity operator might be written,
$\sum_n\!\!\!^{(B)} = \sum_{n=-N}^{N}$,
with the $n=0$ term excluded,
and $n^\dag = -n$.
There is no reason why the microstates could not be relabeled
so that they cover the same range in both cases,
with some convenient code linking the index of a state and its conjugate
in the odd parity case.

For an equilibrium system,
the weight of a microstate is equal to that of its conjugate,
\begin{equation}
w^A_n = w^A_{n^\dag} ,
\end{equation}
whatever the parity of $\hat A$.
This is because an equilibrium system is insensitive to the direction of time,
and hence it is insensitive to the sign of the molecular velocities.
This means that the weight operator is real, $\hat w^* = \hat w$,
as also is the probability operator and the entropy operator.

\subsubsection{Microstate Reversibility}

For an open equilibrium quantum system
(i.e. a sub-system and a reservoir in total isolated),
the sub-system microstate transition
$\{A,n_1\} \stackrel{\tau}{\rightarrow} \{B,n_2\}$
has conjugate
$\{B,n_{2}^{\dag}\} \stackrel{\tau}{\rightarrow}\{A, n_1^{\dag}\}$.
For the present open equilibrium quantum system,
these have equal weight
\begin{equation} \label{Eq:wBA=wABdag}
w^{BA}(n_2,n_1|\tau) = w^{AB}(n_1^\dag,n_2^\dag|\tau) .
\end{equation}
This is called  microstate reversibility.
A detailed justification for it will be given in the
derivation of Eq.~(\ref{Eq:wBA=wABdag2}) in the following section.

The general expression for the macrostate transition weight is
\begin{equation}
w_{\alpha_2\beta_1}^{(2),CD}(\tau)
=
\sum_{k_2,l_1}
w_{\stackrel{\scriptstyle C:\alpha_2 k_2,\alpha_2 k_2}
{D:\beta_1 l_1,\beta_1 l_1}}
^{(2)}(\tau).
\end{equation}
This could be written alternatively as
\begin{equation}
w^{(2)}(\alpha_2^C,\beta_1^D|\tau)
=
\sum_{k_2\in\alpha_2} \sum_{l_1\in\beta_1}
w^{(2)}(\alpha_2^C k_2,\beta_1^D l_1|\tau),
\end{equation}
or as
\begin{equation}
w^{(2)}(\alpha_2^C,\beta_1^D|\tau)
=
\sum_{k_2,l_1}
w^{(2)}
(
\begin{array}{l}
{ C:\alpha_2 k_2,\alpha_2 k_2} \\
{D:\beta_1 l_1,\beta_1 l_1}
\end{array}
|\tau ).
\end{equation}
Recall that the eigenfunction equation for an incomplete operator $\hat C$
is
$\hat C | \zeta^C_{\alpha k} \rangle = C_\alpha | \zeta^C_{\alpha k} \rangle$,
and similarly for the incomplete operator $\hat D$.
In this case macrostate  reversibility is
\begin{eqnarray}
w^{DC}(\beta^\dag,\alpha^\dag|\tau)
& = &
\sum_{k^\dag\in\alpha^\dag} \sum_{l^\dag\in\beta^\dag}
w^{DC}(\beta^\dag l^\dag,\alpha^\dag k^\dag|\tau)
\nonumber \\  & = &
\sum_{k\in\alpha} \sum_{l\in\beta}
w^{DC}(\beta^\dag l^\dag,\alpha^\dag k^\dag|\tau)
\nonumber \\ & = &
\sum_{k\in\alpha} \sum_{l\in\beta}
w^{CD}(\alpha k,\beta l|\tau)
\nonumber \\ & = &
w^{CD}(\alpha,\beta|\tau) .
\end{eqnarray}
In passing to the second equality,
the fact that $k^\dag$ and $l^\dag$ are dummy summation variables
has been used;
for every $k^\dag\in\alpha^\dag$ there is a $k\in\alpha$,
and for every $l^\dag\in\beta^\dag$ there is an $l \in \beta$.
The third equality invokes microstate reversibility.

\subsubsection{Transition Symmetries}

The weight of a sub-system transition
$| \psi_1 \rangle \stackrel{\tau}{\rightarrow} | \psi_2 \rangle$
is the number of adiabatic total trajectories that project onto it.
For each total trajectory,
$| \psi_\mathrm{1s}\psi_\mathrm{1r} \rangle
\stackrel{\tau}{\rightarrow}
| \psi_\mathrm{2s}\psi_\mathrm{2r}  \rangle$
(`s' denotes the sub-system, `r' denotes the reservoir,
allowed wave functions are not entangled),\cite{QSM1}
there are several trajectories related by conjugation and time reversal,
\begin{eqnarray} \label{Eq:8-Traj}
| \psi_\mathrm{1s}\psi_\mathrm{1r} \rangle
\stackrel{\tau}{\rightarrow}
| \psi_\mathrm{2s}\psi_\mathrm{2r}  \rangle
& \;\;\; &
| \psi_\mathrm{2s}\psi_\mathrm{2r} \rangle
\stackrel{-\tau}{\rightarrow}
| \psi_\mathrm{1s}\psi_\mathrm{1r}  \rangle
\nonumber \\
| \psi_\mathrm{1s}^*\psi_\mathrm{1r}^* \rangle
\stackrel{-\tau}{\rightarrow}
| \psi_\mathrm{2s}^*\psi_\mathrm{2r}^*  \rangle
& \;\;\; &
| \psi_\mathrm{2s}^*\psi_\mathrm{2r}^* \rangle
\stackrel{\tau}{\rightarrow}
| \psi_\mathrm{1s}^*\psi_\mathrm{1r}^*  \rangle
\nonumber \\
\langle \psi_\mathrm{1s}\psi_\mathrm{1r} |
\stackrel{-\tau}{\rightarrow}
\langle \psi_\mathrm{2s}\psi_\mathrm{2r}  |
& \;\;\; &
\langle \psi_\mathrm{2s}\psi_\mathrm{2r} |
\stackrel{\tau}{\rightarrow}
\langle \psi_\mathrm{1s}\psi_\mathrm{1r}  |
\nonumber \\
\langle \psi_\mathrm{1s}^*\psi_\mathrm{1r}^* |
\stackrel{\tau}{\rightarrow}
\langle \psi_\mathrm{2s}^*\psi_\mathrm{2r}^*  |
& \;\;\; &
\langle \psi_\mathrm{2s}^*\psi_\mathrm{2r}^* |
\stackrel{-\tau}{\rightarrow}
\langle \psi_\mathrm{1s}^*\psi_\mathrm{1r}^*  | .
\end{eqnarray}
Since the total system is isolated,
these are adiabatic transitions.
When projected onto the sub-system,
all eight trajectories have the same weight
(i.e.\ the same number of trajectories of the total system that involve
$\psi_\mathrm{1s}$ and $\psi_\mathrm{2s}$).

An arbitrary wave function can be expanded in the basis of eigenfunctions
of an arbitrary (complete) operator
\begin{equation}
| \psi \rangle
= \sum_n \psi_n | \zeta_{n} \rangle
, \mbox{ with }
\psi_n =  \langle \zeta_{n} | \psi \rangle.
\end{equation}
Its complex conjugate is
\begin{eqnarray}
| \psi^* \rangle
& = &
\sum_n \psi_n^* | \zeta_{n}^{*} \rangle
\nonumber \\ & = &
\sum_n \psi_n^* | \zeta_{n^\dag} \rangle
\nonumber \\ & = &
\sum_n \psi_{n^\dag}^* | \zeta_{n} \rangle .
\end{eqnarray}
The final equality uses the fact that $n$ is a dummy
summation variable that ranges over all the original and conjugate states.
The parity of the operator that determines the basis
also determines whether or not $n^\dag$ is the same state as $n$.

The probability of a transition,
$| \psi_1 \rangle \stackrel{\tau}{\rightarrow} | \psi_2 \rangle$,
assuming normalized wave functions, is
\begin{eqnarray}
\wp^{(2)}( \psi_2 , \psi_1|\tau)
& = &
\langle \psi_2,\psi_1 |
\hat \wp^{(2)}(\tau)
| \psi_1, \psi_2 \rangle
\\ \nonumber & = &
\sum_{\stackrel{\scriptstyle m_2,n_2}{m_1,n_1}}
 \psi_{2,n_2}^{B*} \psi_{2,m_2}^{B}
\psi_{1,n_1}^{A*} \psi_{1,m_1}^{A}
\wp^{(2)}_{\stackrel{\scriptstyle B:n_2,m_2}{A:n_1,m_1}}(\tau) .
\end{eqnarray}
The probability of a transition,
$| \psi_2^* \rangle \stackrel{\tau}{\rightarrow} | \psi_1^* \rangle$
is
\begin{eqnarray}
\lefteqn{
\wp^{(2)}( \psi_1^* , \psi_2^* |\tau)
}
\\ \nonumber &  = &
\sum_{\stackrel{\scriptstyle m_2,n_2}{m_1,n_1}}
\psi_{1,n_1^\dag}^{A} \psi_{1,m_1^\dag}^{A*}
\psi_{2,n_2^\dag}^{B} \psi_{2,m_2^\dag}^{B*}
\wp^{(2)}_{\stackrel{\scriptstyle A:n_1,m_1}{B:n_2,m_2}}(\tau)
\\ \nonumber &  = &
\sum_{\stackrel{\scriptstyle m_2,n_2}{m_1,n_1}}
\psi_{1,m_1}^{A} \psi_{1,n_1}^{A*}
\psi_{2,m_2}^{B} \psi_{2,n_2}^{B*}
\wp^{(2)}_{\stackrel{\scriptstyle A:m_1^\dag,n_1^\dag}{B:m_2^\dag,n_2^\dag}}
(\tau).
\end{eqnarray}
The probability of a transition,
$| \psi_2 \rangle \stackrel{-\tau}{\rightarrow} | \psi_1 \rangle$
is
\begin{eqnarray}
\lefteqn{
\wp^{(2)}( \psi_1 , \psi_2 |{-\tau})
}
\\ \nonumber &  = &
\sum_{\stackrel{\scriptstyle m_2,n_2}{m_1,n_1}}
\psi_{1,n_1}^{A*} \psi_{1,m_1}^{A}
\psi_{2,n_2}^{B*} \psi_{2,m_2}^{B}
\wp^{(2)}_{\stackrel{\scriptstyle A:n_1,m_1}{B:n_2,m_2}}(-\tau) .
\end{eqnarray}

The reality of the expectation value  means
that the transition probability operator
is Hermitian conjugate,
$\hat \wp^{(2)}(\tau) = \hat \wp^{(2)}(\tau)^\ddag$,
or
\begin{equation}
\wp^{(2)}_{\stackrel{\scriptstyle B:n_2,m_2}{A:n_1,m_1}}(\tau)
=
\wp^{(2)}_{\stackrel{\scriptstyle B:m_2,n_2}{A:m_1,n_1}}(\tau)^* .
\end{equation}

Hence one has three symmetry relations
\begin{equation} \label{Eq:wp2-sym}
\wp^{(2)}_{\stackrel{\scriptstyle n_2,m_2}{n_1,m_1}}(\tau)
=
\wp^{(2)}_{\stackrel{\scriptstyle m_1^\dag,n_1^\dag}{m_2^\dag,n_2^\dag}}
(\tau)
=
\wp^{(2)}_{\stackrel{\scriptstyle n_1,m_1}{n_2,m_2}}(-\tau)
=
\wp^{(2)}_{\stackrel{\scriptstyle m_2,n_2}{m_1,n_1}}(\tau)^*  .
\end{equation}
The operators used for the representation
have not been shown.
The first equality expresses microstate reversibility,
and the second equality expresses time reversibility.
Time reversibility is the combination
of time homogeneity and statistical symmetry.

Note that the operators used for the representation,
here $\hat A$ and $\hat B$,
determine the relationship between the state $n$
and its conjugate, $n^\dag$.
However these symmetry rules  depend upon the probability operator
and not upon the particular operators used for the representation.
Hence one can drop explicit signification of the operators,
with the understanding that a single operator is used for $t_1$
and a single operator, either the same or different, is used for $t_2$.

Recall that this is the \emph{unconditional} transition probability.
For an equilibrium system  time reversibility must hold:
the unconditional probability of observing the forward transition must be equal
to the probability of observing the backward transition,
otherwise there would be a net flux between the states.
This is distinct from the \emph{conditional} transition probability,
which the second law of thermodynamics would ensure was asymmetric in time,
as is further discussed shortly.

For the state transitions the symmetries  mean that
\begin{equation} \label{Eq:wBA=wABdag2}
\wp^{(2),BA}_{n_2n_1}(\tau)
=
\wp^{(2),AB}_{n_1^\dag n_2^\dag}(\tau)
=
\wp^{(2),AB}_{n_1n_2}(-\tau)
=
\wp^{(2),BA}_{n_2n_1}(\tau)^*.
\end{equation}
The first equality here is consistent with,
and is the justification for,
microstate reversibility given above  in terms of weight,
Eq.~(\ref{Eq:wBA=wABdag}).

Again it is emphasized
that these are \emph{unconditional} transition probabilities,
which are related to the conditional  transition probability
by Bayes' theorem,
$\wp^{(2)}(n_2^B,n_1^A|\tau) = \wp^{(2)}(n_2^B|n_1^A,\tau) \wp(n_1^A)$.
As such the time reversibility result,
$\wp^{(2)}(n_2^B,n_1^A|\tau)
= \wp^{(2)}(n_1^A,n_2^B|{-\tau})$ implies that
$\wp^{(2)}(n_2^B|n_1^A,\tau) \wp(n_1^A)
= \wp^{(2)}(n_1^A|n_2^B,-\tau) \wp(n_2^B)$, or
\begin{equation}
\frac{\wp^{(2)}(n_2^B|n_1^A,\tau)}{\wp^{(2)}(n_1^A|n_2^B,-\tau)}
=
\frac{\wp(n_2^B)}{\wp(n_1^A)} .
\end{equation}
That is, the ratio of the  forward and reverse
conditional transition probabilities
is the inverse of the ratio of the probabilities
of the respective starting states.
This is what one would expect from the second law of thermodynamics:
transitions \emph{to} a more probable state are more likely
than are transitions \emph{from} a more probable state.

The transition weight has of course the same symmetries
as the unconditional transition probability.
It was shown above, Eq.~(\ref{Eq:hatUUwp1}),
that the unconditional transition probability operator
could be written in terms of the propagator,
$ \hat \wp^{(2)}(\tau) =
\{ \hat {\cal U}(\tau)^\dag , \hat {\cal U}(\tau)  \hat \wp^{(1)} \}$,
and three other permutations.
This has representation
\begin{equation}
\wp^{(2)}_{\stackrel{\scriptstyle m_2 n_2}{m_1 n_1}}(\tau)
=
\sum_{l_1}
\left< {\cal U}_{m_1n_2}(\tau)^\dag
{\cal U}_{m_2l_1}(\tau) \right>_\mathrm{stoch}
\wp^{(1)}_{l_1 n_1} ,
\end{equation}
with ${\cal U}_{m_1n_2}(\tau)^\dag = {\cal U}_{n_2m_1}(\tau)^*$.
There is no need to explicitly denote the operators used
for the representation.
The symmetry rules applied to this yield
\begin{eqnarray}
\lefteqn{
\sum_{l_1}
\left< {\cal U}_{m_1n_2}(\tau)^\dag
{\cal U}_{m_2l_1}(\tau) \right>_\mathrm{stoch}
\wp^{(1)}_{l_1 n_1}
} \nonumber \\
& = &
\sum_{l_2}
\left< {\cal U}_{n_2^\dag m_1^\dag}(\tau)^\dag
{\cal U}_{n_1^\dag l_2^\dag}(\tau) \right>_\mathrm{stoch}
\wp^{(1)}_{l_2^\dag m_2^\dag}
\nonumber \\ & = &
\sum_{l_2}
\left< {\cal U}_{m_2n_1}(-\tau)^\dag
{\cal U}_{m_1 l_2}(-\tau) \right>_\mathrm{stoch}
\wp^{(1)}_{l_2 n_2}
\nonumber \\ & = &
\sum_{l_1}
\left< {\cal U}_{m_2 n_1 }(\tau)
{\cal U}_{ l_1 n_2}(\tau)^\dag \right>_\mathrm{stoch}
\wp^{(1)}_{l_1 m_1} .
\end{eqnarray}
Of course $\wp^{(1)*}_{l_1 m_1} =\wp^{(1)}_{l_1 m_1} = \wp^{(1)}_{m_1 l_1} $,
and
$\wp^{(1)}_{l_2^\dag m_2^\dag} = \wp^{(1)}_{l_2 m_2}$.

Setting in these  $m_2=n_2$ and summing over $n_2$ one obtains
\begin{eqnarray}
\lefteqn{
\sum_{n_2,l_1}
\left< {\cal U}_{m_1n_2}(\tau)^\dag
{\cal U}_{n_2l_1}(\tau) \right>_\mathrm{stoch}
\wp^{(1)}_{l_1 n_1}
} \nonumber \\
& = &
\sum_{n_2,l_2}
\left< {\cal U}_{n_2^\dag m_1^\dag}(\tau)^\dag
{\cal U}_{n_1^\dag l_2^\dag}(\tau) \right>_\mathrm{stoch}
\wp^{(1)}_{l_2^\dag n_2^\dag}
\nonumber \\ & = &
\sum_{n_2,l_2}
\left< {\cal U}_{n_2n_1}(-\tau)^\dag
{\cal U}_{m_1 l_2}(-\tau) \right>_\mathrm{stoch}
\wp^{(1)}_{l_2 n_2}
\nonumber \\ & = &
\sum_{n_2,l_1}
\left< {\cal U}_{n_2 n_1 }(\tau)
{\cal U}_{ l_1 n_2}(\tau)^\dag \right>_\mathrm{stoch}
\wp^{(1)}_{l_1 m_1} .
\end{eqnarray}

Now
\begin{eqnarray}
{\cal U}_{m^\dag n^\dag}(\tau) & = &
\langle \zeta_{m^\dag} | \hat {\cal U}(\tau) |  \zeta_{n^\dag} \rangle
\nonumber \\ & = &
\langle \zeta_{m}^* | \hat {\cal U}(\tau) |  \zeta_{n}^* \rangle
\nonumber \\ & = &
\langle \zeta_{m} | \hat {\cal U}(\tau)^* |  \zeta_{n} \rangle^*
\nonumber \\ & = &
\left\{ \hat {\cal U}(\tau)^* \right\}_{mn}^*.
\end{eqnarray}
Note that the propagator is neither Hermitian nor of pure parity.
Hence
\begin{eqnarray}
\lefteqn{
{\cal U}_{n_2^\dag m_1^\dag}(\tau)^\dag
{\cal U}_{n_1^\dag l_2^\dag}(\tau)
\wp^{(1)}_{l_2^\dag n_2^\dag}
} \nonumber \\
& = &
{\cal U}_{m_1^\dag n_2^\dag }(\tau)^*
\wp^{(1)}_{n_2 l_2 }
{\cal U}_{l_2^\dag n_1^\dag }(\tau)^\mathrm{T}
\nonumber \\ & = &
\left\{ \hat {\cal U}(\tau)^* \right\}_{m_1n_2}
\wp^{(1)}_{ n_2 l_2}
\left\{ \hat {\cal U}(\tau)^* \right\}_{ l_2 n_1}^\dag .
\end{eqnarray}
After summation over $l_2$ and $n_2$,
this is equivalent to the $\{m_1,n_1\}$ element
of the representation of $\hat {\cal U}(\tau)^*
\hat \wp^{(1)}
\hat {\cal U}(\tau)^\mathrm{T}$.
Hence the symmetry relations may be written for the operators as
\begin{eqnarray}
\left< \hat{\cal U}(\tau)^\dag \hat{\cal U}(\tau) \right>_\mathrm{stoch}
\hat \wp^{(1)}
& = &
\left< \hat {\cal U}(\tau)^*
\hat \wp^{(1)}
\hat {\cal U}(\tau)^\mathrm{T} \right>_\mathrm{stoch}
\nonumber \\ & = &
\left< \hat {\cal U}(-\tau)
\hat \wp^{(1)}
\hat {\cal U}(-\tau)^\dag \right>_\mathrm{stoch}
\nonumber \\ & = &
\hat \wp^{(1)}
\left< \hat {\cal U}(\tau)^\dag
\hat {\cal U}(\tau) \right>_\mathrm{stoch} .
\end{eqnarray}
It has already been established that the propagator is on average unitary,
Eq.~(\ref{Eq:UU=I:proof}).
Hence these become
\begin{eqnarray} \label{Eq:wp1=UUwp1}
\hat \wp^{(1)}
& = &
\left< \hat {\cal U}(\tau)^*
\hat \wp^{(1)}
\hat {\cal U}(\tau)^\mathrm{T} \right>_\mathrm{stoch}
\nonumber \\ & = &
\left< \hat {\cal U}(-\tau)
\hat \wp^{(1)}
\hat {\cal U}(-\tau)^\dag \right>_\mathrm{stoch}
\nonumber \\ & = &
\left< \hat {\cal U}(\tau)
\hat \wp^{(1)}
\hat {\cal U}(\tau)^\dag \right>_\mathrm{stoch}
\nonumber \\ & = &
\left< \hat {\cal U}(-\tau)^*
\hat \wp^{(1)}
\hat {\cal U}(-\tau)^\mathrm{T} \right>_\mathrm{stoch} ,
\end{eqnarray}
the third and fourth equalities following upon taking the complex conjugate
of the first two equalities and using the fact that
the singlet probability operator is real.
The second equality represents
the backward evolution of the probability operator,
(for positive $\tau $).
Changing $\tau$ to $-\tau$
in the second equality provides an alternative derivation
of the third equality,
which represent the forward evolution of the probability operator.
Equating each to the left hand side
shows that the equilibrium probability operator
is on average stationary under the evolution given by
the stochastic dissipative time propagator.
This result has previously been derived using
the reduction condition, Eq.~(\ref{Eq:wp-stationary}).

As has been mentioned,
the unconditional transition probability operator
can also be written in terms of the propagator as
$ \hat \wp^{(2)}(\tau) =
\{\hat {\cal U}(\tau)^\dag \hat \wp^{(1)}, \hat {\cal U}(\tau) \}
= \{\hat {\cal U}(\tau)^\dag  , \hat\wp^{(1)}\hat {\cal U}(\tau) \}$.
These have respective representations
\begin{eqnarray}
\wp^{(2)}_{\stackrel{\scriptstyle m_2 n_2}{m_1 n_1}}(\tau)
& = &
\sum_{l_2}
\left< {\cal U}_{m_1l_2}(\tau)^\dag
\wp^{(1)}_{l_2 n_2}
{\cal U}_{m_2n_1}(\tau) \right>_\mathrm{stoch}
\nonumber \\ & = &
\sum_{l_2}
\left< {\cal U}_{m_1n_2}(\tau)^\dag
\wp^{(1)}_{m_2 l_2}
{\cal U}_{l_2n_1}(\tau) \right>_\mathrm{stoch} .
\end{eqnarray}
Applying the symmetry conditions to this
yields an alternative set of stationary conditions
on the probability operator.
It seems plausible that these should simply
interchange the propagator and its Hermitian conjugate,
yielding
\begin{eqnarray}
\hat \wp^{(1)}
& = &
\left< \hat {\cal U}(\tau)^\mathrm{T}
\hat \wp^{(1)}
\hat {\cal U}(\tau)^* \right>_\mathrm{stoch}
\nonumber \\ & = &
\left< \hat {\cal U}(-\tau)^\dag
\hat \wp^{(1)}
\hat {\cal U}(-\tau) \right>_\mathrm{stoch}
\nonumber \\ & = &
\left< \hat {\cal U}(\tau)^\dag
\hat \wp^{(1)}
\hat {\cal U}(\tau)\right>_\mathrm{stoch}
\nonumber \\ & = &
\left< \hat {\cal U}(-\tau)^\mathrm{T}
\hat \wp^{(1)}
\hat {\cal U}(-\tau)^* \right>_\mathrm{stoch} .
\end{eqnarray}

One can, for example,  take the transpose of the first equality in
Eq.~(\ref{Eq:wp1=UUwp1})
and compare it to the first equality here,
\begin{eqnarray}
\hat \wp^{(1)}
& = &
\left< \hat {\cal U}(\tau)
\hat \wp^{(1)}
\hat {\cal U}(\tau)^\dag \right>_\mathrm{stoch}
\nonumber \\ & = &
\left< \hat {\cal U}(\tau)^\mathrm{T}
\hat \wp^{(1)}
\hat {\cal U}(\tau)^* \right>_\mathrm{stoch} .
\end{eqnarray}
One can see that it is sufficient for
the propagator to be transpose symmetric,
$\hat {\cal U}(\tau)^\mathrm{T} = \hat {\cal U}(\tau)$,
for this to be satisfied.
Similar identities hold for the transpose of each of the
remaining three equalities in Eq.~(\ref{Eq:wp1=UUwp1}).
This does not prove that it is necessary
for the propagator to be transpose symmetric.


In terms of state transitions the uncondition state transition probability is
\begin{equation}
\wp^{(2)}_{n_2,n_1}(\tau)
=
\left< {\cal U}_{n_1n_2}(\tau)^\dag
{\cal U}_{n_2n_1}(\tau) \right>_\mathrm{stoch}
\wp^{(1)}_{n_1 n_1} .
\end{equation}
Hence the symmetry rules yield
\begin{eqnarray} \label{Eq:UUn2-n1}
\lefteqn{
\left<
{\cal U}_{n_2n_1}(\tau)^* \,
{\cal U}_{n_2n_1}(\tau)
\right>_\mathrm{stoch} \wp^{(1)}_{n_1 n_1}
} \nonumber \\
& = &
\left<
{\cal U}_{n_1^\dag n_2^\dag }(\tau)^* \,
{\cal U}_{n_1^\dag n_2^\dag }(\tau)
\right>_\mathrm{stoch} \wp^{(1)}_{n_2^\dag n_2^\dag}
\nonumber \\ & = &
\left<
{\cal U}_{n_1n_2}(-\tau)^* \,
{\cal U}_{n_1n_2}(-\tau)
\right>_\mathrm{stoch} \wp^{(1)}_{n_2 n_2}
\nonumber \\ & = &
\left<
{\cal U}_{n_2n_1}(\tau) \,
{\cal U}_{n_2n_1}(\tau)^*
\right>_\mathrm{stoch} \wp^{(1)}_{n_1 n_1} .
\end{eqnarray}
Of course $\wp^{(1)}_{n_2^\dag n_2^\dag} = \wp^{(1)}_{n_2 n_2}$.

Assume that the operator $\hat A$ has pure parity,
$\varepsilon_A = \pm 1$,
then  $\hat A^* = \varepsilon_A \hat A$,
and $ A_{n^\dag} = \varepsilon_A A_n$,
and similarly for the operator $\hat B$.
In this case the time correlation function has the symmetries
\begin{equation}
C_{BA}(\tau)  = C_{AB}(-\tau)  = C_{A^*B^*}(\tau)
= \varepsilon_A  \varepsilon_B  C_{AB}(\tau).
\end{equation}
In terms of the propagator these are
\begin{eqnarray}
C_{BA}(\tau) & = &
\sum_{n_1,n_2}
\left< {\cal U}_{n_1n_2}^{AB}(\tau)^\dag
{\cal U}_{n_2n_1}^{BA}(\tau) \right>_\mathrm{stoch}
B_{n_2} A_{n_1} \wp^{A}_{n_1} ,
\nonumber \\ & = &
\sum_{n_1,n_2}
\left< {\cal U}_{n_1 n_2 }^{BA}(-\tau)^\dag
{\cal U}_{n_2 n_1 }^{AB}(-\tau) \right>_\mathrm{stoch}
\nonumber \\ && \mbox{ } \times
B_{n_1} A_{n_2} \wp^{B}_{n_1}
\nonumber \\ & = &
\sum_{n_1,n_2}
\left< {\cal U}_{n_1 n_2 }^{BA}(\tau)^\dag
{\cal U}_{n_2 n_1 }^{AB}(\tau) \right>_\mathrm{stoch}
B_{n_1^\dag} A_{n_2^\dag} \wp^{B}_{n_1}
\nonumber \\ & = &
\varepsilon_A  \varepsilon_B  \sum_{n_1,n_2}
\left< {\cal U}_{n_1 n_2 }^{BA}(\tau)^\dag
{\cal U}_{n_2 n_1 }^{AB}(\tau) \right>_\mathrm{stoch}
\nonumber \\ && \mbox{ } \times
B_{n_1} A_{n_2} \wp^{B}_{n_1}.
\end{eqnarray}
Here $\wp^{B}_{n_1} \equiv \wp^{(1),BB}_{n_1n_1}$ etc.
Hence
\begin{eqnarray}
\lefteqn{
\left< {\cal U}_{n_1 n_2 }^{BA}(-\tau)^\dag
{\cal U}_{n_2 n_1 }^{AB}(-\tau) \right>_\mathrm{stoch}
} \nonumber \\
& =&
\varepsilon_A  \varepsilon_B
\left< {\cal U}_{n_1 n_2 }^{BA}(\tau)^\dag
{\cal U}_{n_2 n_1 }^{AB}(\tau) \right>_\mathrm{stoch}
\nonumber \\ & =&
\left< {\cal U}_{n_1^\dag n_2^\dag }^{BA}(\tau)^\dag
{\cal U}_{n_2^\dag n_1^\dag }^{AB}(\tau) \right>_\mathrm{stoch} .
\end{eqnarray}
The equality of the left hand side with the second right hand side here
is equivalent to the equality of the second and third right hand sides
in Eq.~(\ref{Eq:UUn2-n1}).

\subsection{Symmetries of the Time Propagator}

The goal of this section is to elucidate
the symmetries or properties of the most likely  time propagator
that are a consequence of microscopic reversibility
(as opposed to the properties of the average of the product
of the time propagators analyzed above).

The above results for the consequences of microstate reversibility
for the transition probability yielded two main results
for the time propagator:
it was on average unitary,
$ \langle \hat{{\cal U}}(\tau)^\dag \hat{{\cal U}}(\tau)
\rangle_\mathrm{stoch}
= \langle \hat{{\cal U}}(\tau) \, \hat{{\cal U}}(\tau)^\dag
\rangle_\mathrm{stoch}
= \hat{\mathrm I}$,
and evolving under its action the probability distribution
was on average stationary,
$ \hat \wp
=\langle \hat{{\cal U}}(\tau) \hat \wp \, \hat{{\cal U}}(\tau)^\dag
\rangle_\mathrm{stoch}
= \langle \hat{{\cal U}}(\tau)^\dag \hat \wp \, \hat{{\cal U}}(\tau)
\rangle_\mathrm{stoch}$.

Both of these results invoke
the average of the product of the time propagator.
In terms of the time propagator itself,
one can define the most likely value as
\begin{equation}
\hat{\overline{\cal U}}(\tau)
=
\left< \hat{\cal U}(\tau) \right>_\mathrm{stoch} .
\end{equation}
This assumes a Gaussian distribution for the stochastic propagator
(means equal modes),
which is reasonable on physical grounds.

It does \emph{not} follow from the unitary condition
that the Hermitian conjugate is the inverse,
$ \hat{\overline{\cal U}}(\tau)^\dag
\ne \hat{\overline{\cal U}}(\tau)^{-1}$.
The reason is that the unitary condition
applies only to the average of the product,
which is not equal to the product of the averages.
In general terms, the inverse of an operator is only defined
in conjunction with its product with the original operator,
and so one cannot expect $\hat{\overline{\cal U}}(\tau)^{-1}$
to have any physical meaning.
Instead, it is $\hat{\overline{\cal U}}(-\tau)$
that plays the role of undoing the action of the original propagator,
the meaning of which must be carefully understood.
(In contrast, in the adiabatic case,
$\hat U^0(-\tau) = \hat U^0(\tau)^{-1} = \hat U^0(\tau)^{\dag}$.)

By definition, the most likely wave function
at time $t_1+\tau$ given that the system has wave function
$\psi_1$ at time $t_1$ is
\begin{equation}
|  \psi_2 \rangle
\equiv
| \overline \psi(\tau|\psi_1) \rangle
=
\hat{\overline{\cal U}}(\tau)
|  \psi_1 \rangle .
\end{equation}
In Eq.~(\ref{Eq:8-Traj}) above,
eight statistically equivalent trajectories were given.
Focussing on the left hand column projected onto the sub-system,
(i.e.\ trajectories that originate from $\psi_\mathrm{s1}$ or its conjugate),
the most likely trajectories are
\begin{eqnarray} \label{Eq:4-Traj}
| \psi_2 \rangle = \hat{\overline{\cal U}}(\tau) | \psi_1  \rangle
&   &
| \psi_2^* \rangle = \hat{\overline{\cal U}}(-\tau) | \psi_1^*  \rangle
\nonumber \\
\langle \psi_2 | =
\langle \psi_1 | \hat{\overline{\cal U}}(-\tau)^\mathrm{T}
& \mbox{and} &
 \langle \psi_2^* | =
\langle \psi_1^* | \hat{\overline{\cal U}}(\tau)^\mathrm{T} .
\end{eqnarray}
(Although all eight trajectories in Eq.~(\ref{Eq:8-Traj})
have the same weight,
due to thermodynamic irreversibility,
the most likely trajectory obtained by maximizing over $\psi_2$
for fixed $\psi_1$ is not necessarily related to
the most likely trajectory obtained by maximizing over $\psi_1$
for fixed $\psi_2$. This is why only the four trajectories
starting at $\psi_1$ are compared here.)
The second and third equations must be equivalent
to the complex conjugate and to the Hermitian conjugate
of the first equation respectively,
and the fourth equation must be equivalent to applying both operations
simultaneously to the first equation.
These mean that
\begin{equation} \label{Eq:olU(t)*=olU(-t)}
\hat{\overline{\cal U}}(\tau)^* = \hat{\overline{\cal U}}(-\tau)
,\mbox{ and }
\hat{\overline{\cal U}}(\tau)^\dag =
\hat{\overline{\cal U}}(-\tau)^\mathrm{T}  .
\end{equation}
These are equivalent to each other.
The fourth equation yields the identity $\hat{\overline{\cal U}}(\tau)^{\dag*}
= \hat{\overline{\cal U}}(\tau)^\mathrm{T}$.
(From these, one cannot conclude anything about
the transpose symmetry of  the most likely propagator.)

The physical interpretation of this symmetry,
$\hat{\overline{\cal U}}(-\tau)^* = \hat{\overline{\cal U}}(\tau)$,
and the reason why it is consistent with thermodynamic irreversibility,
is as follows.
In general if $\psi_2 =  \overline \psi(\tau|\psi_1 )$,
then $\psi_2$ has higher entropy then $\psi_1$.
But it does not follow that it is also
the most likely origin of $\psi_1$,
$\psi_2 \ne  \overline \psi(-\tau|\psi_1 )$,
because the adiabatic contribution is odd in $\tau$.
(It is only the reservoir contribution that can be expected to be even).
However, $\psi_2^*$ also has higher entropy then $\psi_1^*$.
So in this case one can expect
$\psi_2^*$ to be the most likely origin of $\psi_1^*$,
which is to say that it is
the most likely backward transition from $\psi_1^*$,
$\psi_2^* =  \overline \psi(-\tau|\psi_1^*)$.
This satisfies both the adiabatic part and the thermodynamic part.

Mathematically,
for small time intervals one expects to be able to write
the most likely propagator
as the sum of an adiabatic part and a reservoir part,
$\hat{\overline{\cal U}}(\tau) =  \hat{{\cal U}}^0(\tau)
+ \hat{\overline{\cal U}}\,^\mathrm{r}(\tau)$.
This expectation is borne out by the results in Paper II.\cite{QSM2}
Hence the most likely propagator is
\begin{eqnarray}
\hat{\overline {\cal U}}(\tau)  & = &
\hat{\mathrm I}
+ \frac{\tau}{i\hbar} \hat{\cal H}
+ |\tau| \hat{{\cal U}}\,^\mathrm{r}
+ {\cal O}(\tau^2).
\end{eqnarray}
The absolute value of the time interval arises because
for the present equilibrium case,
the reservoir should be insensitive to the direction of time.
By inspection,
the adiabatic part has the requisite symmetry,
\begin{equation}
\hat{{\cal U}}^0(-\tau)^* = \hat{{\cal U}}^0(\tau).
\end{equation}
The most likely reservoir part by design
has the symmetry $\hat{\overline{\cal U}}\,^\mathrm{r}(-\tau) =
\hat{\overline{\cal U}}\,^\mathrm{r}(\tau)$.
Since the reservoir is insensitive to the sign of the velocities,
the propagator arising from it must also be real,
\begin{equation}
\hat{\overline{\cal U}}\,^\mathrm{r}(\tau)^* =
\hat{\overline{\cal U}}\,^\mathrm{r}(\tau) .
\end{equation}
Combining these, the most likely propagator must have the symmetry
deduced above,
\begin{equation}
\hat{\overline{\cal U}}(-\tau)^* = \hat{\overline{\cal U}}(\tau) .
\end{equation}
Although this third derivation has been obtained explicitly
in the small time interval limit,
it would not strain credibility to suppose that it holds in general.
Indeed, the first two arguments were not restricted to any particular
time regime.

\comment{ 
In the adiabatic case,
$\hat{{\cal U}}^0(\tau)^{-1} = \hat{{\cal U}}^0(-\tau)$,
and one can use these interchangeably.
In the open case,
one has to deal with
$\hat{\overline{\cal U}}(-\tau)$ rather than
$\hat{\overline{\cal U}}(\tau)^{-1}$
because the reciprocal only has meaning in a product,
and one has to take the average of the product
and not the product of the averages.
} 

                \section{Fluctuation Form of the Second Entropy}
\label{Sec:S2-flucn}
\setcounter{equation}{0}

\subsection{Fluctuation Form of the First Entropy}

The expectation value of the first entropy is
\begin{eqnarray} \label{Eq:S1-flucn}
S^{(1)<>}(\psi)
& = &
\frac{ \langle   \psi  | \hat S^{(1)} |  \psi   \rangle
}{N(\psi)}
\nonumber \\ & = &
\overline S\,^{(1)}
+ \langle   \Delta \psi  | \hat S''|  \Delta \psi   \rangle .
\end{eqnarray}
Here the departure from the most likely wave function is
$\Delta \psi = \psi - \overline  \psi$.
The most likely wave function is an eigenfunction
of the entropy operator,
\begin{equation}
\hat S^{(1)} |  \overline \psi   \rangle
= S_0 |  \overline \psi   \rangle
= \frac{-E_0}{T} |  \overline \psi   \rangle ,
\end{equation}
The final equality holding in the canonical equilibrium case,
$\hat S^{(1)} = -\hat {\cal H} /T$.
The entropy fluctuation operator is
\begin{equation}
\hat S'' =
\frac{1}{N(\overline \psi)}
\left[ \hat S^{(1)} - S_0 \hat{\mathrm I} \right].
\end{equation}
Hence $\hat S'' |  \overline \psi   \rangle = |  0   \rangle$.
The most likely wave function may be taken to be normalized,
$N(\overline \psi)=1$.

In the canonical equilibrium case,
the most likely wave function has adiabatic time dependence\cite{QSM2}
\begin{equation}
|  \overline \psi(t)   \rangle
=
e^{E_0t/i\hbar}  |  \overline \psi(0)   \rangle .
\end{equation}
It will be shown below that at least to linear order in the time step,
and almost certainly in general,
this adiabatic evolution is the complete evolution.

\subsection{Second Entropy}

The expectation value of the second entropy
for the wave state transition $\psi_1 \stackrel{\tau}{\rightarrow} \psi_2$
is
\begin{equation}
S^{(2)<>}(\psi_2,\psi_1|\tau)
=
\frac{ \langle  \psi_2 ,  \psi_1  |
\hat S^{(2)}(\tau)
|  \psi_1  ,  \psi_2  \rangle
}{N(\psi_2) N(\psi_1)}.
\end{equation}

With $\overline \psi$ being the most likely equilibrium state,
one can define
\begin{equation}
\hat a(\tau) \equiv
\left.
\frac{\partial^2 S^{(2)<>}(\psi_2,\psi_1|\tau)
}{\partial \langle \psi_2 | \, \partial | \psi_2 \rangle
}\right|_{\overline \psi} ,
\end{equation}
\begin{equation}
\hat c(\tau) \equiv
\left.
\frac{\partial^2 S^{(2)<>}(\psi_2,\psi_1|\tau)
}{ \partial \langle \psi_1  | \, \partial | \psi_1 \rangle
}\right|_{\overline \psi} ,
\end{equation}
\begin{equation}
\hat b(\tau) \equiv
\left.
\frac{\partial^2 S^{(2)<>}(\psi_2,\psi_1|\tau)
}{  \partial \langle \psi_2 |  \, \partial | \psi_1 \rangle
}\right|_{\overline \psi} .
\end{equation}
These derivative are evaluated at $\psi_1  = \overline \psi$,
and $ \psi_2 = \overline \psi(\tau)$.

As mentioned above,
the most likely wave state has an adiabatic evolution,
$\overline \psi(t) = e^{E_0t/i\hbar}  \overline \psi(0) $,
and so the departures from the most likely equilibrium states
are $\Delta \psi_1 = \psi_1 - \overline \psi$
and $\Delta \psi_2 = \psi_2 - \overline \psi(\tau)$.
(It remains to show that the adiabatic evolution
of the most likely wave function
equals the most likely evolution.
This explicit result for the time dependence
of the  most likely wave function is not required for
most of what follows.)
The expectation value of the second entropy may be expanded
to quadratic order about the most likely values,
\begin{eqnarray}
\lefteqn{
S^{(2)<>}(\psi_2,\psi_1|\tau)
}  \\ \nonumber
& \approx &
\overline S\,^{(1)}
+ \langle \Delta \psi_2 | \hat a(\tau) |  \Delta \psi_2  \rangle
+
\langle \Delta \psi_1 | \hat c(\tau) | \Delta \psi_1  \rangle
 \\ \nonumber && \mbox{ }
+
\langle \Delta \psi_2 | \hat b(\tau) |  \Delta \psi_1 \rangle
+
\langle \Delta \psi_1 | \hat b(\tau)^\dag |  \Delta \psi_2 \rangle .
\end{eqnarray}
The constant term comes from the reduction condition.

The three symmetries are
statistical symmetry combined with time homogeneity,
\begin{equation}
S^{(2)}( \psi_2, \psi_1|\tau)
 =
S^{(2)}( \psi_1, \psi_2|{-\tau}) ,
\end{equation}
microscopic reversibility,
\begin{equation}
S^{(2)}( \psi_2, \psi_1|\tau) =
S^{(2)}( \psi_1^*, \psi_2^*|\tau) ,
\end{equation}
and reality,
\begin{equation}
S^{(2)}( \psi_2, \psi_1|\tau) =
S^{(2)}( \psi_2, \psi_1|\tau)^* .
\end{equation}

Statistical symmetry is
\begin{equation}
S^{(2)}( \{\psi_2, t_2\}; \{\psi_1,t_1\}) =
S^{(2)}( \{\psi_1, t_1\}; \{\psi_2,t_2\}) ,
\end{equation}
and  time homogeneity is
\begin{equation}
S^{(2)}( \{\psi_2, t_2\}; \{\psi_1,t_1\}) =
S^{(2)}( \psi_2, \psi_1|t_{21}) .
\end{equation}
It is the combination of these that gives the above symmetry.

Statistical symmetry implies that
\begin{equation}
\hat a(-\tau) = \hat c(\tau)
,\mbox{ and }
\hat b(-\tau)
= \hat b(\tau)^\dag .
\end{equation}
Reality implies that
\begin{equation}
\hat a(\tau)^\dag = \hat a(\tau)
,\mbox{ and }
\hat c(\tau)^\dag = \hat c(\tau) .
\end{equation}
The form of the two cross terms in the second entropy guarantees reality
for the contribution from these two terms.
Microscopic reversibility implies that
\begin{equation}
\hat a(\tau)^* = \hat c(\tau)
,\mbox{ and }
\hat b(\tau)^*   = \hat b(\tau)^\dag .
\end{equation}
Hence $\hat a(\tau) = \hat a(\tau)^\dag = \hat a(-\tau)^* $,
and $\hat b(\tau) = \hat b(\tau)^{*\dag} = \hat b(-\tau)^* $.

It ought to be emphasized
that the fluctuation form for the expectation value
of the second entropy is an approximation.
Essentially, the exact second entropy operator
is a four dimensional object,
whereas the fluctuation form
consists of the sum of four  operators (reduced to three by symmetry)
that are each two dimensional objects.

\subsection{Most Likely Terminus}

The derivative of the fluctuation form for the second entropy with respect to
the terminal arrival departure  $\left< \Delta \psi_2 \right| $ is
\begin{equation}
\frac{ \partial S^{(2)<>}( \psi_2, \psi_1;\tau)
}{\partial \left< \Delta \psi_2 \right| }
 =
\hat a(\tau) \left| \Delta \psi_2 \right>
+
\hat b(\tau)  \left| \Delta \psi_1 \right> .
\end{equation}
(The unusual phrase `terminal arrival departure'
means the  terminus at the arrival end of the transition (i.e.\ the $\psi_2$)
expressed as the departure from the most likely value
(i.e.\  $\psi_2 - \overline \psi(t_2)$).)
Setting this to zero,
the most likely transition departure,
$\Delta \overline \psi_2
\equiv \overline \psi(\tau|\psi_1) -   \overline \psi(\tau)$,
is
\begin{equation}
| \Delta \overline \psi_2 \rangle
=
- \hat a(\tau)^{-1} \hat b(\tau) |  \Delta \psi_1 \rangle .
\end{equation}

Since in quantum mechanics one is dealing with operator equations
that are linear homogeneous functions of the wave function,
this implies that the most likely destination itself is
\begin{equation}
| \overline \psi(\tau|\psi_1) \rangle
=
- \hat a(\tau)^{-1} \hat b(\tau) |  \psi_1 \rangle  .
\end{equation}
Consequently,
the most likely wave function trajectory must satisfy
\begin{equation}
| \overline \psi(\tau) \rangle
=
- \hat a(\tau)^{-1} \hat b(\tau) |  \overline \psi \rangle  .
\end{equation}
The meaning of this latter result,
and its consistency with the assertion made above
that the evolution of the most likely wave function is adiabatic,
will be discussed below.
(See also the appendix.)

\subsection{Reduction Condition}

The second entropy may be re-written in terms of the departure
from the most likely terminus,
$ \psi_2 -  \overline \psi_2$,
\begin{eqnarray}
\lefteqn{
S^{(2)<>}( \psi_2, \psi_1|\tau)
= }
\\ \nonumber & &
\overline S^{(1)}
+
\left< \Delta \psi_2 - \Delta \overline \psi_2 \right|
\hat a(\tau)
\left| \Delta \psi_2 - \Delta \overline \psi_2 \right>
\\  \nonumber && \mbox{ }
+
\left< \Delta \psi_1 \right|
\left\{ \hat c(\tau)
- \hat b(\tau)^\dag \hat a(\tau)^{-1}  \hat b(\tau) \right\}
\left| \Delta \psi_1\right> .
\end{eqnarray}

The reduction condition is that in the most likely state,
the second entropy reduces to the first entropy,\cite{NETDSM,AttardII}
$S^{(2)<>}(\overline \psi_2, \psi_1 |\tau)
=
S^{(1)<>}( \psi_1)$,
(see also Eq.~(\ref{Eq:S1n=S2nm-red})).
In view of the fluctuation expression for the first entropy,
Eq.~(\ref{Eq:S1-flucn}),
this implies that
\begin{equation}
\hat c(\tau)
- \hat b(\tau)^\dag \hat a(\tau)^{-1}  \hat b(\tau)
=
\hat S'' .
\end{equation}
This result must hold for each value of the time step $\tau$.

\subsection{Small Time Expansion}

Almost all of \S IIIA4 of Paper II\cite{QSM2} goes through unchanged,
giving the small time expansions as
\begin{equation}
\hat a(\tau) =
\frac{ - 1}{|\tau|} \hat \lambda^{-1}
+ \hat a_{0} + \widehat \tau \hat a_{0}' + {\cal O}(\tau) ,
\end{equation}
\begin{equation}
\hat b(\tau) =
\frac{1}{|\tau|} \hat \lambda^{-1}
+ \hat b_{0} + \widehat \tau \hat b_{0}' + {\cal O}(\tau),
\end{equation}
and
\begin{equation}
\hat c(\tau) =
\frac{-1}{|\tau|} \hat \lambda^{-1}
+ \hat a_{0} - \widehat \tau \hat a_{0}' + {\cal O}(\tau) .
\end{equation}
The operator $\hat \lambda$ is real, symmetric
(and hence it is Hermitian), and positive definite.

From the symmetries given above,
$\hat a(\tau) = \hat a(\tau)^\dag = \hat a(-\tau)^* $,
and $ \hat a(-\tau) = \hat c(\tau)$,
one can see that
the unprimed $\hat a$ are real and self-adjoint
and equal the unprimed $\hat c$,
and the primed $\hat a$ are imaginary  and self-adjoint
and equal the negative of the primed $\hat c$.
Also,
since $\hat b(\tau) = \hat b(\tau)^{*\dag} = \hat b(-\tau)^* $,
the unprimed $\hat b$ are real and self-adjoint,
and the primed $\hat b$ are imaginary and anti-self-adjoint.

The reduction condition yields
\begin{equation}
\hat a_{0} + \hat b_{0}
= \frac{1}{2 } \hat S'' .
\end{equation}
The is the same as in Paper II\cite{QSM2}.
This result is the analogue of the result for classical fluctuations
in macrostates or microstates given in Ref.~\onlinecite{NETDSM}.

Using this, the expansion
for the most likely terminal wave function  becomes
\begin{eqnarray}
|  \overline \psi_2 \rangle
& = &
- \hat a(\tau)^{-1} \hat b(\tau)  | \psi_1 \rangle
\nonumber \\  & = &
\left[ \hat{\mathrm I}
+ |\tau| \hat \lambda \hat a_{0}
+  \tau \hat \lambda \hat a_{0}'
\right]
\nonumber \\ &  & \mbox{ } \times
\left[
\hat{\mathrm I} +
|\tau| \hat \lambda \hat b_{0} +  \tau  \hat \lambda \hat b_{0}'
\right] |  \psi_1 \rangle
\nonumber \\ & = &
\left|  \psi_1 \right>
+
\tau  \hat \lambda \left[ \hat a_{0}' + \hat b_{0}' \right] | \psi_1 \rangle
\nonumber \\ &  & \mbox{ }
+ \frac{ |\tau| }{2 } \hat \lambda \hat S''
|  \psi_1 \rangle
+ {\cal O}(\tau^2) .
\end{eqnarray}

The adiabatic evolution must be contained in the reversible term,
which is the one that is proportional to $\tau$,
$| \dot \psi_1^0 \rangle
= (1/i\hbar) \hat {\cal H}  | \psi_1 \rangle$.
There may be reversible reservoir contributions,
but since the reservoir is nothing but a perturbation
to the adiabatic evolution,
they can be neglected compared to the reversible adiabatic term.
One cannot neglect the irreversible reservoir  term
because there is not an adiabatic irreversible term with which to compare it.
Since it is the only irreversible term,
and since irreversibility is an essential ingredient in the evolution
that follows directly from the second law of thermodynamics,
one must retain  the irreversible reservoir  term.

Hence equating the reversible term here with the adiabatic evolution
of the wave function, one must have
\begin{equation} 
\hat \lambda \left[ \hat a_{0}' + \hat b_{0}' \right]
=
\frac{1}{i\hbar} \hat {\cal H}   .
\end{equation}
Both sides are imaginary.
Since the right hand side is anti-Hermitian,
one must have
\begin{equation} 
\hat \lambda \left[ \hat a_{0}' + \hat b_{0}' \right]
=
- \left[ \hat a_{0}'^\dag + \hat b_{0}'^\dag \right] \hat \lambda^\dag
=
- \left[ \hat a_{0}' - \hat b_{0}'  \right] \hat \lambda .
\end{equation}
Contrary to what was asserted following Eq.~(3.32) of Paper II,\cite{QSM2}
this does not imply that  $\hat b_{0}' = \hat 0$.
Fortunately, that erroneous assertion does not affect
anything in that paper beyond the assertion itself.

Inserting this adiabatic result,
the formula for the most likely transition terminal is
\begin{eqnarray}
| \overline \psi(\tau|\psi_1) \rangle
& = &
|  \psi_1 \rangle
+ \frac{ \tau }{i\hbar}  \hat{\cal H} |  \psi_1 \rangle
+ \frac{|\tau|}{2} \hat \lambda  \hat S'' |  \psi_1 \rangle  .
\end{eqnarray}

As mentioned above,
in order for this linear homogeneous form
to result from the formula for the most likely departure
of the transition terminal,
the most likely wave state has to cancel both sides.
The validity of doing this can be seen by inserting into this result
the most likely wave function as the initial wave function,
which gives
\begin{eqnarray}
| \overline \psi(\tau|\overline \psi) \rangle
& = &
|  \overline \psi \rangle
+ \frac{ \tau }{i\hbar}  \hat{\cal H} |  \overline \psi  \rangle
+ \frac{|\tau|}{2} \hat \lambda  \hat S'' | \overline \psi  \rangle
\nonumber \\ & = &
\left[ 1 + \frac{ \tau E_0 }{i\hbar}  \right] |  \overline \psi  \rangle .
\end{eqnarray}
This is just the adiabatic development
of the most likely wave function to linear order in the time step,
as foreshadowed above.
That to leading the reservoir does not contribute to this evolution
is to be expected because by definition there is no entropy gradient
at this most likely wave state and so there is no
driving force from the reservoir.

Several points follow from this.
The final argument regarding the vanishing of the
entropy gradient holds for all equilibrium systems,
not just the canonical one.
It is therefore plausible that in the general equilibrium case
evolution of the most likely wave function should be reversible
and adiabatic.
This appears to hold beyond the leading order in the time step.
This is reasonable and it says that there is a unique
most likely trajectory, $\overline \psi(t)$.

From the above result for the most likely terminal wave function
of a transition,
one can identify the most likely propagator as
\begin{equation}
\hat{\overline{\cal U}}(\tau)
= \hat{\mathrm I}
+ \frac{ \tau }{i\hbar}  \hat{\cal H}
+ \frac{|\tau|}{2} \hat \lambda  \hat S''
+{\cal O}(\tau^2) .
\end{equation}
It consists of an adiabatic part, the first two terms,
and a dissipative part,
the final term, which is proportional to the gradient of the reservoir entropy.
The fluctuation operator $\hat \lambda $
may also be called the drag or friction operator.

The most likely propagator must satisfy the reversibility
condition obtained above, Eq.~(\ref{Eq:olU(t)*=olU(-t)}),
\begin{equation}
\hat{\overline{\cal U}}(\tau)^* = \hat{\overline{\cal U}}(-\tau)
,\mbox{ and }
\hat{\overline{\cal U}}(\tau)^\dag =
\hat{\overline{\cal U}}(-\tau)^\mathrm{T}  .
\end{equation}
Since both $\hat \lambda$ and $ \hat S''$  are real,
one sees that this is indeed satisfied.

\subsection{Stochastic, Dissipative Propagator}

To the most likely propagator may be added a stochastic operator
to give the stochastic dissipative time propagator
\begin{equation}
\hat{{\cal U}}(\tau)
=
\hat{\overline{\cal U}}(\tau) + \hat{\cal R} .
\end{equation}

The probability distribution of the  stochastic operator
must obey certain properties that are derived from the symmetry
conditions on the transition probability
(equivalently transition entropy)
and the other conditions that were established above for the propagator.

It was established above that the propagator is on average unitary,
Eq.~(\ref{Eq:UU=I:proof}).
From this it follows that the variance of the
stochastic part of the propagator must be
\begin{equation} \label{Eq:<RR>=lambda S}
\left< \hat{\cal R}^\dag \hat{\cal R}
\right>_\mathrm{stoch}
=
\frac{-|\tau|}{2}
\left[ \hat \lambda  \hat S''
+  \hat S'' \hat \lambda  \right] .
\end{equation}

The symmetry relations showed that
the equilibrium probability distribution was stationary
under the action of the stochastic propagator,
Eq.~(\ref{Eq:wp1=UUwp1}),
\begin{equation}
\hat \wp^{(1)}
=
\left< \hat {\cal U}(\tau)
\hat \wp^{(1)}
\hat {\cal U}(\tau)^\dag \right>_\mathrm{stoch} .
\end{equation}
For the case of a canonical equilibrium system,
the Maxwell-Boltzmann probability operator
is readily shown to be stationary to under the adiabatic evolution,
\begin{equation}
\hat \wp^{(1)}_\mathrm{MB}
=
\left< \hat {\cal U}^0(\tau)
\hat \wp^{(1)}_\mathrm{MB}
\hat {\cal U}^0(\tau)^\dag \right>_\mathrm{stoch} .
\end{equation}
(Hint: The Hamiltonian operator is proportional to the entropy operator
and so commutes with it.)
Assuming that in general the equilibrium probability distribution
is stationary under the adiabatic evolution,
then the stationary condition becomes
\begin{equation}
\left< \hat{\cal R}^\dag \hat \wp^{(1)} \hat{\cal R}
\right>_\mathrm{stoch}
=
\frac{-|\tau|}{2}
\left[ \hat \lambda  \hat S'' \hat \wp^{(1)}
+  \hat \wp^{(1)} \hat S'' \hat \lambda  \right] .
\end{equation}
This may be called the quantum fluctuation dissipation theorem.

If the probability operator and the time propagator commute,
then the quantum fluctuation dissipation theorem
is identical to the unitary condition.

If the entropy operator  $\hat S^{(1)}$  or $ \hat S''$
and the fluctuation operator $\hat \lambda$ commute,
and if the entropy operator and the stochastic operator $\hat{\cal R}$
commute,
then the quantum fluctuation dissipation theorem
is identical to the unitary condition.

\subsection{Ansatz for the Drag and Stochastic Operators}

The drag and stochastic operators represent the
contribution of the reservoir to the evolution
of the sub-system wave function.
The concept of a reservoir is a key element in statistical mechanics
that allows detailed analysis and calculation
to be focussed on that part of the system
that is of immediate interest while treating the interactions
with the neighboring and far regions
in a   averaged, integrated fashion.
The reservoir contributions
represent a perturbation on the adiabatic motion of the sub-system
that has to obey certain statistical and symmetry rules
(the unitary and stationarity conditions deduced above)
but are otherwise arbitrary.

Perhaps the simplest ansatz for the  drag and stochastic operators
is to construct them from the entropy eigenfunctions such that
they are diagonal in the entropy representation.
The drag operator can be taken to be of the form
\begin{equation}
\hat \lambda
=
\sum_{\alpha,g} \lambda_\alpha
 | \zeta^S_{\alpha g} \rangle  \langle \zeta^S_{\alpha g} | .
\end{equation}
The coefficients can be chosen as desired,
although since $\hat \lambda $ must be Hermitian and positive semi-definite,
$\lambda_\alpha  = \lambda_\alpha ^* \ge 0$.
(If the drag coefficient vanishes, then
the corresponding  stochastic coefficient vanishes ---see below.)
The stochastic operator  may similarly be taken to be
\begin{equation}
\hat {\cal R}
=
\sum_{\alpha ,g} r_\alpha
| \zeta^S_{\alpha g} \rangle  \langle \zeta^S_{\alpha g} | .
\end{equation}

Inserting these into the unitary condition,
Eq.~(\ref{Eq:<RR>=lambda S}) one obtains
\begin{equation}
\left< r_\alpha ^* r_\beta \right>_\mathrm{stoch}
=
-\delta_{\alpha \beta} |\tau| ( S_\alpha  - S_0 ) \lambda_\alpha  .
\end{equation}
This says that the coefficients of the stochastic operator
of different entropy states are uncorrelated.

With this ansatz,
there is no stochastic contribution
to the evolution of the ground state, $r_0 = 0$.
The ground state drag coefficient may be non-zero, $\lambda_0 > 0$,
but the dissipative contribution to the ground state evolution
is zero, $ ( S_n - S_0 ) \lambda_n = 0$ if $n=0$.
With this ansatz, the evolution of the ground state
is purely adiabatic.

Since this ansatz for the drag operator
is diagonal in the entropy representation,
it commutes with the entropy operator,
$ \hat \lambda \hat S = \hat S  \hat \lambda$,
and with the entropy fluctuation operator,
$ \hat \lambda \hat S'' = \hat S'' \hat \lambda$.
(Incidently, this means that with this ansatz
the propagator is transpose symmetric,
$\hat {\cal U}(\tau)^\mathrm{T} = \hat {\cal U}(\tau)$.)
For the same reason the stochastic operator
commutes with the entropy operator,
$ \hat {\cal R} \hat S = \hat S  \hat {\cal R}$.
Since the probability operator is the  exponential the entropy operator,
it commutes with the drag operator, the stochastic operator,
and the  the entropy fluctuation operator.
Hence it commutes with the stochastic dissipative propagator,
$ \hat {\cal U}(\tau) \hat \wp = \hat \wp  \,\hat {\cal U}(\tau)$.
Hence with this ansatz,
the stationarity condition is automatically satisfied
if the unitary condition is satisfied,
\begin{equation}
\left< \hat {\cal U}(\tau)^\dag \hat \wp \, \hat {\cal U}(\tau)
\right>_\mathrm{stoch}
=
\left< \hat {\cal U}(\tau)^\dag   \hat {\cal U}(\tau)
\right>_\mathrm{stoch} \hat \wp
= \hat \wp.
\end{equation}

    \section{Nature of Non-Equilibrium Probability}   \label{Sec:neQSM}
\setcounter{equation}{0}

Non-equilibrium systems are defined
by the increase in total entropy with time.
Hence in setting out the nature of non-equilibrium probability,
the functions given in \S \ref{Sec:Nat-Prob}
become time dependent.
For example the weight operator is $\hat w(t)$.

The relationships between weight, probability, and entropy
are  unchanged functionally.
The microstate weight is
\begin{equation}
w_n^A(t) = \langle \zeta_n^A | \hat w(t) | \zeta_n^A \rangle,
\end{equation}
the macrostate weight is
\begin{equation}
w_\alpha^C(t) =
\sum_{k\in\alpha} \!\!^{(C)} w_{\alpha k}^C(t)
=
\sum_{k\in\alpha} \!\!^{(C)}
\langle \zeta_{\alpha k}^C | \hat w(t) | \zeta_{\alpha k}^C \rangle,
\end{equation}
and the total weight is
\begin{equation}
W(t)
= \sum_\alpha \!\!^{(C)} w_\alpha^C(t)
=
\sum_{\alpha,k} \!\!^{(C)} w_{\alpha k}^C(t) .
\end{equation}
Similarly the non-equilibrium probability operator
is $\hat \wp(t) = \hat w(t) /W(t)$,
and the non-equilibrium entropy operator is
$\hat S(t) = k_\mathrm{B} \ln  \hat w(t)  $.
The distinction between the expectation value of the entropy,
$S^{<>}_{mn}(t) \equiv \langle \zeta_m| \hat S(t) |\zeta_n \rangle$
and the entropy of states
$S_{mn}(t) \equiv
k_\mathrm{B} \ln \langle \zeta_m| \hat w(t) |\zeta_n \rangle$
remains,
with
$\wp_{mn}(t) =  e^{S_{mn}(t)/k_\mathrm{B} } /W(t)$.

In the case of transitions (c.f.\ \S \ref{Sec:trans}),
some changes are required.
Consider the transition from the state $n_1^A$ at $t_1$
to $n_2^B$ at $t_2 = t_1 + \tau$,
and define the mid-time as $t \equiv [t_2 +t_1]/2$.
The weight of this transition may be denoted
$w^{(2)}(n_1^A,t_1;n_2^B,t_2) \equiv w^{(2)}(n_1^A,n_2^B|\tau,t) $.
Unlike the equilibrium case, one cannot assume time homogeneity,
and so both time arguments have to appear in the
transition operator, $\hat w(\tau,t) $.
Practically all of the analysis of \S \ref{Sec:trans}
goes through with the change $\tau \Rightarrow \tau,t$.

One significant change in the non-equilibrium case
is  the reduction condition.
This now takes the form
\begin{equation}
\sum_{n_2^B} w^{(2)}(n_1^A,t_1;n_2^B,t_2) =
w_{n_1^A}(t_1) \sqrt{ \frac{W(t_2)}{W(t_1)} } ,
\end{equation}
and
\begin{equation}
\sum_{n_1^A} w^{(2)}(n_1^A,t_1;n_2^B,t_2) =
w_{n_2^B}(t_2)\sqrt{ \frac{W(t_1)}{W(t_2)} } .
\end{equation}
The  scale factor is new compared to the equilibrium case
and is necessary for the correct symmetry for the total transition weight,
as is now shown
(see also \S8.3.1 of Ref.~[\onlinecite{NETDSM}]).
The total transition weight,
$W^{(2)}(t_2,t_1) \equiv W^{(2)}(\tau,t)$, is
\begin{eqnarray}
W^{(2)}(t_2,t_1)
& = & \sum_{n_2^B,n_1^A} w^{(2)}(n_1^A,t_1;n_2^B,t_2)
\nonumber \\ & = &
 \sum_{n_1^A} w_{n_1^A}(t_1) \sqrt{ \frac{W(t_2)}{W(t_1)} }
\nonumber \\ & = &
\sqrt{ W(t_2) W(t_1) } .
\end{eqnarray}
The same result follows if the order of the summations is changed.
Because of this scaling,
the reduction condition on the weight operators is
\begin{equation}
\mbox{Tr}^{(1)}_1 \hat w^{(2)}(t_2,t_1)
= \sqrt{ \frac{W(t_1)}{W(t_2)} } \hat w(t_2) ,
\end{equation}
and
\begin{equation}
\mbox{Tr}^{(1)}_2 \hat w^{(2)}(t_2,t_1)
= \sqrt{ \frac{W(t_2)}{W(t_1)} } \hat w(t_1) ,
\end{equation}
where the subscript on the trace indicates which time is traced over.
In terms of the transition probability operator,
$\hat \wp^{(2)}(t_2,t_1) = \hat w^{(2)}(t_2,t_1)/W^{(2)}(t_2,t_1)$,
the reduction condition is
\begin{eqnarray}
\mbox{Tr}^{(1)}_1  \hat \wp^{(2)}(t_2,t_1)
& = &
\frac{1}{\sqrt{ W(t_2) W(t_1) }}
\mbox{Tr}^{(1)}_1  \hat w^{(2)}(t_2,t_1)
\nonumber \\ & = &
\frac{1}{\sqrt{ W(t_2) W(t_1) }}
\sqrt{ \frac{W(t_1)}{W(t_2)} }
\hat w(t_2)
\nonumber \\ & = &
\hat \wp(t_2) ,
\end{eqnarray}
and, analogously,
$\mbox{Tr}^{(1)}_2 \hat \wp^{(2)}(t_2,t_1) = \hat \wp(t_1)$.

The propagator gives the evolution of the wave function
in the non-equilibrium system,
\begin{equation}
| \psi(t_2) \rangle = \hat {\cal U}(t_2,t_1)  | \psi(t_1) \rangle .
\end{equation}
It may also be written $\hat {\cal U}(\tau,t)$.
The  conditional transition probability operator
is a two-time operator
that can be written   as the composition of the two one-time time propagators,
\begin{equation}
\hat \wp^{(2),\mathrm{cond}}(t_2,t_1)
=
\left< \left\{
\hat {\cal U}(t_2,t_1)^\dag ,  \hat {\cal U}(t_2,t_1)
\right\} \right>_\mathrm{stoch}  .
\end{equation}
Accordingly
the unconditional transition probability operator
is the composition
of the conditional transition probability operator
and the singlet probability operator
that can be arranged in four ways,
\begin{eqnarray}
\lefteqn{
\hat \wp^{(2)}(t_2,t_1)
} \nonumber \\
& = &
\left< \left\{
\hat {\cal U}(t_2,t_1)^\dag ,  \hat {\cal U}(t_2,t_1) \hat \wp^{(1)}(t_1)
\right\} \right>_\mathrm{stoch}
\nonumber \\ & = &
\left< \left\{
\hat \wp^{(1)}(t_1) \hat {\cal U}(t_2,t_1)^\dag ,  \hat {\cal U}(t_2,t_1)
\right\} \right>_\mathrm{stoch}
\nonumber \\ & = &
\left< \left\{
\hat {\cal U}(t_2,t_1)^\dag , \hat \wp^{(1)}(t_2) \hat {\cal U}(t_2,t_1)
\right\} \right>_\mathrm{stoch}
\nonumber \\ & = &
\left< \left\{
\hat {\cal U}(t_2,t_1)^\dag \hat \wp^{(1)}(t_2) ,  \hat {\cal U}(t_2,t_1)
\right\} \right>_\mathrm{stoch} .
\end{eqnarray}
Taking the traces of this and using the reduction condition,
one obtains the stationarity condition and the unitary condition
for the propagator.




\appendix
    \section{Fluctuations Revisited}  
\setcounter{equation}{0}

\renewcommand{\theequation}{A.\arabic{equation}}

\subsection{Most Likely Sub-Space}

In the text, particularly  \S\ref{Sec:S2-flucn},
the fluctuation forms of the first and second entropy were analyzed.
This analysis was based upon the departure from
the most likely wave function
$\Delta \psi(t) \equiv \psi - \overline \psi(t)$.
In this appendix, a clearer, more rigorous, and more general
analysis of fluctuations is given,
as well as some discussion of the consequences of the analysis.

Denote the eigenstates of maximum entropy by the principle quantum number 0.
Taking into account degeneracy,
the corresponding  eigenfunctions are $\zeta^S_{0g}$,
$g=1,2,\ldots$
with $\hat S | \zeta^S_{0g}   \rangle =
S_0 |  \zeta^S_{0g}  \rangle$.
(The total entropy operator and the reservoir entropy operator
are equivalent and are both denoted here by $\hat S$.)
For the present equilibrium system,
the sub-space is constant in time.
One can ignore the time dependence of the phase angle
of the eigenfunctions.
The sub-space that these span may be called the most likely sub-space.
It can also be called the entropy ground state,
since it is the space with the  entropy eigenvalue
that is smallest in magnitude,
(equivalently, since the eigenvalues are negative,
the largest signed eigenvalue).
The projection operator for this most likely sub-space is
\begin{equation}
\hat{\cal P}_0 \equiv \sum_g
|\zeta^S_{0g}\rangle \langle \zeta^S_{0g} | .
\end{equation}
The projector for the orthogonal sub-space, the excited sub-space, is
$\hat{\cal P}_{\!\!\perp} \equiv \hat{\mathrm I} - \hat{\cal P}_0  $.

In general an operator can be decomposed
into its projections onto the two sub-spaces,
\begin{eqnarray}
\hat O & = &
[ \hat{\cal P}_0  + \hat{\cal P}_{\!\!\perp} ]
\hat O
[ \hat{\cal P}_0  + \hat{\cal P}_{\!\!\perp} ]
\nonumber \\ & = &
\hat{\cal P}_0  \hat O \hat{\cal P}_0
+
\hat{\cal P}_0  \hat O  \hat{\cal P}_{\!\!\perp}
+
\hat{\cal P}_{\!\!\perp} \hat O \hat{\cal P}_0
+
\hat{\cal P}_{\!\!\perp} \hat O \hat{\cal P}_{\!\!\perp}
\nonumber \\ & \equiv &
\hat O_{00} + \hat O_{0\perp} + \hat O_{\!\perp 0} + \hat O_{\!\perp\perp} .
\end{eqnarray}

The projection of the wave function $\psi(t)$
onto the ground state sub-space is
\begin{eqnarray}
|\psi_0(t)\rangle
& \equiv &
\hat{\cal P}_0 | \psi(t)  \rangle
\nonumber \\ & = &
\sum_g \langle \zeta^S_{0g} | \psi(t)  \rangle \,
|\zeta^S_{0g}\rangle .
\end{eqnarray}
One could also write $ \overline \psi(t) \equiv \psi_0(t)  $.
The fluctuation of a wave state is the part orthogonal to the
most likely sub-space,
\begin{equation}
| \Delta \psi(t)  \rangle
\equiv \hat{\cal P}_{\!\!\perp} | \psi(t)  \rangle.
\end{equation}

With $\psi_0 \equiv \overline \psi$
and $ \psi_\perp \equiv \Delta \psi$ ,
the first entropy in fluctuation form is
\begin{eqnarray}
S^{<>}(\psi) & = &
\frac{\langle \psi | \hat S | \psi \rangle }{\langle \psi | \psi \rangle }
\nonumber \\ & = &
\frac{\langle \psi | \hat S_{00} + \hat S_{\perp\perp}| \psi \rangle
}{
N(\overline \psi) + N(\Delta \psi) }
\nonumber \\ & = &
\frac{ S_0 N(\overline \psi)
+ \langle \Delta \psi | \hat S_{\perp\perp}| \Delta \psi \rangle
}{
N(\overline \psi) + N(\Delta \psi) }
\nonumber \\ & = &
S_0  +
\frac{1}{N(\overline \psi)}
\langle \Delta \psi | \hat S_{\perp\perp} - S_0 \hat{\mathrm I}_{\perp\perp}
| \Delta \psi \rangle
\nonumber \\ &  & \mbox{ }
+ {\cal O} (\Delta \psi^3).
\end{eqnarray}
Hence
\begin{equation}
\hat S''_{\perp\perp} \equiv
\frac{1}{N(\overline \psi)}
\left[ \hat S_{\perp\perp} - S_0 \hat{\mathrm I}_{\perp\perp} \right] .
\end{equation}
One can take $N(\overline \psi) = 1$.
(Actually it is best \emph{not} to include
the magnitude in the definition of the fluctuation operator.)

The first entropy fluctuation operator just obtained
is confined to the orthogonal sub-space,
$\hat S'' = \hat S''_{\perp\perp}$.
Similarly,
the fluctuation operators for the second entropy defined
in the text, $\hat a(\tau)$, $\hat b(\tau)$, and $\hat c(\tau)$,
are confined to the orthogonal sub-space.

Maximizing the fluctuation form for the second entropy
yields as usual
\begin{eqnarray}
| \Delta \overline \psi_2 \rangle
& = &
- \hat a(\tau)^{-1} \hat b(\tau)  | \Delta \psi_1  \rangle
\nonumber \\ & = &
\hat{ \overline{\cal U}}_{\!\perp\perp}(\tau)  | \Delta \psi_1  \rangle .
\end{eqnarray}
Notice that the most likely propagator that is
defined here is confined to the sub-space
orthogonal to the most likely sub-space,
$\hat{ \overline{\cal U}}_{\!\perp\perp}(\tau)
\equiv
\hat{\cal P}_{\!\!\perp}  \hat{ \overline{\cal U}}(\tau)
\hat{\cal P}_{\!\!\perp}$.

The orthogonal part of the most likely propagator is explicitly
\begin{equation}
\hat{ \overline{\cal U}}_{\!\perp\perp}(\tau)
=
\hat{\mathrm I}_{\!\perp\perp}
+ \frac{\tau}{i\hbar}  \hat{\cal H}_{\!\perp\perp}
+ \frac{|\tau|}{2} \hat \lambda_{\!\perp\perp} \hat S''_{\!\perp\perp} ,
\end{equation}
where $\hat{\mathrm I}_{\!\perp\perp} = \hat{\cal P}_{\!\!\perp}$.
Only the orthogonal projection of the drag operator
contributes to this, $\hat \lambda _{\!\perp\perp}$,
because the fluctuation entropy operator is
originally defined on the orthogonal sub-space.
The projection of the fluctuation operator is
\begin{equation}
\hat S''_{\!\perp\perp}
= \hat S_{\!\perp\perp}  - S_0 \hat{\cal P}_{\!\!\perp}  ,
\end{equation}
and the continuation of this to the full Hilbert space is
$\hat S'' = \hat S  - S_0 \hat{\mathrm I} $.
The entropy operator is block diagonal and may be written
\begin{eqnarray}
\hat S
& = &
\hat S_{00} + \hat S_{\!\perp\perp}
\nonumber \\ & = &
S_0 \hat{\cal P}_{0} + \hat S''_{\!\perp\perp}
+ S_0 \hat{\cal P}_{\!\!\perp}
\nonumber \\ & = &
S_0 \hat{\mathrm I} + \hat S''_{\!\perp\perp} .
\end{eqnarray}
Because the entropy operator is block diagonal,
one can write
\begin{eqnarray}
 \hat \lambda_{\!\perp\perp} \hat S''_{\!\perp\perp}
& = &
\hat \lambda_{\!\perp 0} \hat S''_{0\perp}
+
\hat \lambda_{\!\perp\perp} \hat S''_{\!\perp\perp}
\nonumber \\ & = &
\left\{ \hat \lambda \hat S'' \right\}_{\!\perp\perp}
\nonumber \\ & = &
\left\{ \hat \lambda [ \hat S - S_0 \hat{\mathrm I} ]
\right\}_{\!\perp\perp} .
\end{eqnarray}
One has an analogous result for
$ \hat S''_{\!\perp\perp} \hat \lambda_{\!\perp\perp} $.

For future reference,
In the canonical equilibrium case,
the energy operator is proportional to the entropy operator,
$ \hat{\cal H} = -T \hat S$,
and so it is similarly block diagonal,
$ \hat{\cal H} =\hat{\cal H}_{00} + \hat{\cal H}_{\!\perp\perp}$.

Accordingly, all of the terms on the right hand side
of the propagator on the orthogonal sub-space
are of the form of orthogonal projections of operators
on the full Hermitian space.
Hence it does not strain credulity to write for the
most likely part of the propagator on the full Hermitian space
\begin{equation}
\hat{\overline{\cal U}}(\tau)
=
\hat{\mathrm I}
+ \frac{\tau}{i\hbar}  \hat{\cal H}
+ \frac{|\tau|}{2} \hat \lambda \left[ \hat S - S_0 \hat{\mathrm I} \right].
\end{equation}

Adding a stochastic operator,
the full stochastic dissipative propagator is
\begin{equation}
\hat{\cal U}(\tau)
=
\hat{\mathrm I}
+ \frac{\tau}{i\hbar}  \hat{\cal H}
+ \frac{|\tau|}{2} \hat \lambda \left[ \hat S - S_0 \hat{\mathrm I} \right]
+ \hat{\cal R} .
\end{equation}

The unitary condition to linear order in the time step is
\begin{eqnarray}
\hat{\mathrm I} & = &
\left< \hat{\cal U}(\tau)^\dag \hat{\cal U}(\tau) \right>_\mathrm{stoch}
\nonumber \\ & = &
\sum_{\alpha,\beta,\gamma= 0,\perp}
\left< \hat{\cal U}_{\alpha\beta}(\tau)^\dag
\hat{\cal U}_{\beta\gamma}(\tau) \right>_\mathrm{stoch}
\nonumber \\ & = &
\hat{\mathrm I}
+
\frac{|\tau|}{2}
\left[  \hat \lambda_{0 \perp} + \hat \lambda_{\perp \perp}  \right]
\left[ \hat S_{\perp \perp} - S_0 \hat{\mathrm I}_{\perp \perp} \right]
\nonumber \\ &  & \mbox{ }
+
\frac{|\tau|}{2}
\left[ \hat S_{\perp \perp} - S_0 \hat{\mathrm I}_{\perp \perp} \right]
\left[ \hat \lambda_{\perp 0 }  + \hat \lambda_{\perp \perp}  \right]
\nonumber \\ &  & \mbox{ }
+ \sum_{\alpha,\beta,\gamma= 0,\perp}
\left< \hat{\cal R}_{\alpha\beta}^\dag
\hat{\cal R}_{\beta\gamma} \right>_\mathrm{stoch} .
\end{eqnarray}
The third equality uses the fact that
$ \hat S - S_0 \hat{\mathrm I} =
\hat S_{\perp \perp} - S_0 \hat{\mathrm I}_{\perp \perp} $,
since $\hat S_{0 0} = S_0 \hat{\mathrm I}_{0 0}$.
By definition the average of the stochastic operator vanishes,
$\hat{\overline{\cal R}} = \langle \hat{\cal R} \rangle_\mathrm{stoch} = 0$.

Explicitly this gives for the $00$ component
\begin{equation}
0
=
\langle\hat{\cal R}_{0\perp }^\dag
\hat{\cal R}_{\perp 0 } \rangle_\mathrm{stoch}
+
\langle\hat{\cal R}_{00}^\dag
\hat{\cal R}_{00} \rangle_\mathrm{stoch} .
\end{equation}
For the $0\!\perp$ component it gives
\begin{eqnarray}
\lefteqn{
\frac{-|\tau|}{2}
\hat \lambda_{0\perp}
\{ \hat S_{\perp\perp} - S_0 \hat{\mathrm I}_{\perp\perp} \}
} \nonumber \\
& =  &
\langle\hat{\cal R}_{0\perp }^\dag
\hat{\cal R}_{\perp \perp } \rangle_\mathrm{stoch}
+
\langle\hat{\cal R}_{00}^\dag
\hat{\cal R}_{0\perp} \rangle_\mathrm{stoch} .
\end{eqnarray}
For the $\perp\! 0$ component it gives
\begin{eqnarray}
\lefteqn{
\frac{-|\tau|}{2}
\{ \hat S_{\perp\perp} - S_0 \hat{\mathrm I}_{\perp\perp} \}
\hat \lambda_{\perp 0}
} \nonumber \\
& =  &
\langle\hat{\cal R}_{\perp 0}^\dag
\hat{\cal R}_{0 0 } \rangle_\mathrm{stoch}
+
\langle\hat{\cal R}_{\perp\perp}^\dag
\hat{\cal R}_{\perp 0} \rangle_\mathrm{stoch} .
\end{eqnarray}
For the $\perp \perp$ component it gives
\begin{eqnarray}
\frac{-|\tau|}{2}
\left[ \hat \lambda_{\perp \perp}
\{ \hat S_{\perp\perp} - S_0 \hat{\mathrm I}_{\perp\perp} \}
+
\{ \hat S_{\perp\perp} - S_0 \hat{\mathrm I}_{\perp\perp} \}
\hat \lambda_{\perp \perp} \right]
\nonumber \\ 
=
\langle\hat{\cal R}_{\perp 0}^\dag
\hat{\cal R}_{0 \perp} \rangle_\mathrm{stoch}
+
\langle\hat{\cal R}_{\perp\perp}^\dag
\hat{\cal R}_{\perp \perp} \rangle_\mathrm{stoch} . \hspace{1cm}
\end{eqnarray}
From the $00$ component one concludes that
$\hat{\cal R}_{0 0 } =  \hat{\cal R}_{\perp 0 } = 0$.
(One would also get $\hat{\cal R}_{0 \perp} = 0$
from the unitary condition in the form
$\langle \hat{\cal U}(\tau)\hat{\cal U}(\tau)^\dag  \rangle_\mathrm{stoch}
= \hat{\mathrm I}$.)
Combining this result with the $0\!\perp$ and the  $\perp\! 0$  components
implies that
$\hat\lambda_{0 \perp} = \hat\lambda_{\perp 0 }  = 0$.
Hence only the orthogonal components of the stochastic and dissipative
operators are non-zero,
and these are related by
\begin{eqnarray}
\lefteqn{
\langle\hat{\cal R}_{\perp\perp}^\dag
\hat{\cal R}_{\perp \perp} \rangle_\mathrm{stoch}
} \nonumber \\ & = &
\frac{-|\tau|}{2}
\left[ \hat \lambda_{\perp \perp} \hat S_{\perp\perp}
+  \hat S_{\perp\perp} \hat \lambda_{\perp \perp}
-2 S_0 \hat \lambda_{\perp \perp} \right] .
\end{eqnarray}
One concludes that for the above propagator,
there is no stochastic or dissipative coupling between
the ground state and the excited states,
and there is no stochastic or dissipative term in the ground state.

This conclusion is reason to argue
that the above form for the propagator is not completely satisfactory.
In essence the problem arises because
the dissipative driving force acts only on the orthogonal sub-space
\begin{eqnarray}
\hat S - S_0 \hat{\mathrm I}
& = &
\hat S_{\perp \perp} - S_0 \hat{\mathrm I}_{\perp \perp} ,
\end{eqnarray}
since $\hat S_{0 0} = S_0 \hat{\mathrm I}_{0 0}$.
The consequence of this is that the ground state is decoupled
from the excited states.
For example, the norm of the fluctuation does not decay
but is on average constant,
$\langle N(\Delta \psi(t)) \rangle_\mathrm{stoch}
= \langle N(\Delta \psi(0)) \rangle_\mathrm{stoch}$,
which is counter-intuitive.

The present derivation only gives explicitly
the evolution on the orthogonal sub-space.
It does not preclude additional terms in the full propagator of the form
$ \hat{\Lambda}_{00}(\tau)$,
$ \hat{\Lambda}_{0\perp}(\tau)$,
and
$ \hat{\Lambda}_{\perp 0}(\tau)$,
which would circumvent the decoupling.

\subsection{Cross-Coupling Ansatz}

In view of the last comment, one possibility is
\begin{equation}
\hat{\cal U}(\tau)
=
\hat{\mathrm I}
+ \frac{\tau}{i\hbar}  \hat{\cal H}
+ \frac{|\tau|}{2} \hat \lambda \left[ \hat S - S_0 \hat{\cal P}_\perp \right]
+ \hat{\cal R} .
\end{equation}
The  projection of this on the orthogonal sub-space is unchanged from above,
and it therefore in this case follows from the derivation.
But it is emphasized that there is nothing in
the present derivation that justifies this form for the propagator
for the ground state.
And it is also worth pointing out
that in the classical case the dissipative force
vanishes on the most likely trajectory,\cite{NETDSM}
just as it would vanish in the present quantum case
except for this alternative formulation.

Inserting this into the unitary condition
in the form
$\langle \hat{\cal U}(\tau)^\dag \hat{\cal U}(\tau) \rangle_\mathrm{stoch}
= \hat{\mathrm I}$
leads to
\begin{equation}
\frac{-|\tau|}{2}
\left[ \hat \lambda \hat S +  \hat S \hat \lambda
-  S_0 \hat \lambda  \hat{\cal P}_\perp
-  S_0 \hat{\cal P}_\perp  \hat \lambda \right]
=
\langle\hat{\cal R}^\dag \hat{\cal R} \rangle_\mathrm{stoch}  .
\end{equation}

The $00$ component of this is
\begin{eqnarray}
\lefteqn{
\frac{-|\tau|}{2}
\{  \hat \lambda_{00} \hat S_{00} + \hat S_{00} \hat \lambda_{00} \}
} \nonumber \\
& =  &
-|\tau|  S_0 \hat \lambda_{00}
\nonumber \\ & =  &
\langle\hat{\cal R}_{0\perp }^\dag
\hat{\cal R}_{\perp 0 } \rangle_\mathrm{stoch}
+
\langle\hat{\cal R}_{00}^\dag
\hat{\cal R}_{00} \rangle_\mathrm{stoch} .
\end{eqnarray}
The $0\!\perp$ component is
\begin{eqnarray}
\lefteqn{
\frac{-|\tau|}{2}
\left[
\hat \lambda_{0\perp}
\{ \hat S_{\perp\perp} - S_0 \hat{\mathrm I}_{\perp\perp} \}
+
\hat S_{00} \hat \lambda_{0\perp}
\right]
} \nonumber \\
& =  &
\frac{-|\tau|}{2} \hat \lambda_{0\perp} \hat S_{\perp\perp}
\nonumber \\ & =  &
\langle\hat{\cal R}_{0\perp }^\dag
\hat{\cal R}_{\perp \perp } \rangle_\mathrm{stoch}
+
\langle\hat{\cal R}_{00}^\dag
\hat{\cal R}_{0\perp} \rangle_\mathrm{stoch} .
\end{eqnarray}
The $\perp\! 0$ component is
\begin{eqnarray}
\lefteqn{
\frac{-|\tau|}{2}
\left[
\{ \hat S_{\perp\perp} - S_0 \hat{\mathrm I}_{\perp\perp} \}
\hat \lambda_{\perp 0}
+
\hat \lambda_{\perp 0} \hat S_{00}\right]
} \nonumber \\
& =  &
\frac{-|\tau|}{2} \hat S_{\perp\perp} \hat \lambda_{\perp0}
\nonumber \\ & =  &
\langle\hat{\cal R}_{\perp 0}^\dag
\hat{\cal R}_{0 0 } \rangle_\mathrm{stoch}
+
\langle\hat{\cal R}_{\perp\perp}^\dag
\hat{\cal R}_{\perp 0} \rangle_\mathrm{stoch} .
\end{eqnarray}
For the $\perp \perp$ component it gives
\begin{eqnarray}
\frac{-|\tau|}{2}
\left[ \hat \lambda_{\perp \perp}
\{ \hat S_{\perp\perp} - S_0 \hat{\mathrm I}_{\perp\perp} \}
+
\{ \hat S_{\perp\perp} - S_0 \hat{\mathrm I}_{\perp\perp} \}
\hat \lambda_{\perp \perp} \right]
\nonumber \\ 
=
\langle\hat{\cal R}_{\perp 0}^\dag
\hat{\cal R}_{0 \perp} \rangle_\mathrm{stoch}
+
\langle\hat{\cal R}_{\perp\perp}^\dag
\hat{\cal R}_{\perp \perp} \rangle_\mathrm{stoch} . \hspace{1cm}
\end{eqnarray}
One obtains variants on these from the unitary condition
in the form
$\langle \hat{\cal U}(\tau)\,\hat{\cal U}(\tau)^\dag  \rangle_\mathrm{stoch}
= \hat{\mathrm I}$.
It is possible to satisfy these with consistent expressions for the drag
and stochastic operators that couple the ground and excited states.

The feasibility of such a coupling can be explored more explicitly
with a specific ansatz for the drag and stochastic operators.
The drag operator can be taken to be of the form
\begin{eqnarray}
\hat \lambda
& = &
\sum_{n,g,h} \lambda_n  | \zeta^S_{ng} \rangle  \langle \zeta^S_{nh} |
\\ \nonumber & & \mbox{ }
+
\sum_{n>0,g,h}
\left\{  \lambda_n' | \zeta^S_{0h} \rangle  \langle \zeta^S_{ng} |
+
 \lambda_n'^* | \zeta^S_{ng} \rangle  \langle \zeta^S_{0h} |
\right\} .
\end{eqnarray}
Since $\hat \lambda $ must be Hermitian and positive semi-definite,
$\lambda_n = \lambda_n^* \ge 0$.
(The drag coefficient vanishes if and only if
the corresponding  stochastic coefficient vanishes.)
The stochastic operator  may be taken to be of the form
\begin{eqnarray}
\hat {\cal R}
& = &
\sum_{n,g,h} r_n  | \zeta^S_{ng} \rangle  \langle \zeta^S_{nh} |
\\ \nonumber & & \mbox{ }
+
\sum_{n>0,g,h}
\left\{  r_n' | \zeta^S_{0h} \rangle  \langle \zeta^S_{ng} |
+
r_n'' | \zeta^S_{ng} \rangle  \langle \zeta^S_{0h} |
\right\} .
\end{eqnarray}

The results that follow can be simplified somewhat by
taking $g=h$ in these expressions for $\hat \lambda$ and $\hat R$.
In this case the number of degenerate entropy states, $N_n^\mathrm{S}$,
is replaced by unity in the following formulae.

The left hand side of the unitary condition is
\begin{eqnarray}
\lefteqn{
\frac{-|\tau|}{2}
\left[ \hat \lambda \hat S +  \hat S \hat \lambda
-  S_0 \hat \lambda  \hat{\cal P}_\perp
-  S_0 \hat{\cal P}_\perp  \hat \lambda \right]
} \nonumber \\  & = &
-|\tau|
\sum_{n,g,h} [ S_n -  \delta_{n0}^\dag S_0 ] \lambda_n
| \zeta^S_{ng} \rangle  \langle \zeta^S_{nh} |
\nonumber \\  &  & \mbox{ }
\frac{-|\tau|}{2}
\sum_{n>0,g,h}
\left\{
 \lambda_n' [ S_n - \delta_{n0}^\dag S_0] \,
| \zeta^S_{0h} \rangle  \langle \zeta^S_{ng} |
\right. \nonumber \\ && \left. \mbox{ }
+
\lambda_n'^* [ S_n - \delta_{n0}^\dag S_0] \,
| \zeta^S_{ng} \rangle  \langle \zeta^S_{0h} \right\}
\end{eqnarray}
where $\delta_{n0}^\dag \equiv 1-\delta_{n0}$.

The right hand side of the unitary condition is
\begin{eqnarray}
\lefteqn{
\langle\hat{\cal R}^\dag \hat{\cal R} \rangle_\mathrm{stoch}
} \\ \nonumber & = &
\sum_{n;g,h} \sum_{m;f,k}
\langle r_n^* r_m \rangle_\mathrm{stoch}
 | \zeta^S_{nh} \rangle  \langle \zeta^S_{ng} |
 | \zeta^S_{mf} \rangle  \langle \zeta^S_{mk} |
\\ \nonumber &  & \mbox{ }   
+
\sum_{n;g,h} \sum_{m>0,f,k}
\\ \nonumber &  & \mbox{ }
\left\{
\langle r_n^*  r_m' \rangle_\mathrm{stoch}
| \zeta^S_{nh} \rangle  \langle \zeta^S_{ng} |
| \zeta^S_{0f} \rangle  \langle \zeta^S_{mk} |
\right. \\ \nonumber &  & \left. \mbox{ }
+
\langle r_n^*  r_m'' \rangle_\mathrm{stoch}
| \zeta^S_{nh} \rangle  \langle \zeta^S_{ng} |
| \zeta^S_{mk} \rangle  \langle \zeta^S_{0f} |
\right\}
\\ \nonumber &  & \mbox{ }   
+
\sum_{n>0,g,h} \sum_{m;f,k}
\\ \nonumber &  & \mbox{ }
\left\{
\langle  r_n'^* r_m \rangle_\mathrm{stoch}
| \zeta^S_{ng} \rangle  \langle \zeta^S_{0h} |
| \zeta^S_{mf} \rangle  \langle \zeta^S_{mk} |
\right. \\ \nonumber &  & \left.  \mbox{ }
+
\langle  r_n''^* r_m \rangle_\mathrm{stoch}
| \zeta^S_{0h} \rangle  \langle \zeta^S_{ng} |
| \zeta^S_{mf} \rangle  \langle \zeta^S_{mk} |
\right\}
\\ \nonumber &  & \mbox{ }  
+
\sum_{n>0,g,h} \sum_{m>0;f,k}
\\ \nonumber &  & \mbox{ }
\left<
\left\{  r_n'^* |  \zeta^S_{ng}  \rangle \langle \zeta^S_{0h}  |
+
r_n''^* |   \zeta^S_{0h} \rangle \langle \zeta^S_{ng} | \right\}
\right. \\ \nonumber &  & \left. \mbox{ }
\left\{  r_m' | \zeta^S_{0k} \rangle  \langle \zeta^S_{mf} |
+
r_m'' | \zeta^S_{mf} \rangle  \langle \zeta^S_{0k} | \right\}
\right>_\mathrm{stoch}
\\ \nonumber & = & 
\sum_{n;g,h} N_n^\mathrm{S}
\langle r_n^* r_n \rangle_\mathrm{stoch}
 | \zeta^S_{ng} \rangle  \langle \zeta^S_{nh} |             
\\ \nonumber &  & \mbox{ } 
+
\sum_{m>0,h,k} N_0^\mathrm{S}
\langle r_0^*  r_m' \rangle_\mathrm{stoch}
| \zeta^S_{0h} \rangle    \langle \zeta^S_{mk} |
\\ \nonumber &  & \mbox{ }
+
\sum_{m>0,h,f} N_m^\mathrm{S}
\langle r_m^*  r_m'' \rangle_\mathrm{stoch}
| \zeta^S_{mh} \rangle   \langle \zeta^S_{0f} |                 
\\ \nonumber &  & \mbox{ } 
+
\sum_{n>0,g,k} N_0^\mathrm{S}
\langle r_n'^* r_0 \rangle_\mathrm{stoch}
| \zeta^S_{ng} \rangle    \langle \zeta^S_{0k} |
\\ \nonumber &  & \mbox{ }
+
\sum_{n>0,k,h} N_n^\mathrm{S}
\langle r_n''^* r_n \rangle_\mathrm{stoch}
| \zeta^S_{0h} \rangle   \langle \zeta^S_{nk} |  
\\ \nonumber &  & \mbox{ }  
+
\sum_{n>0,g} \sum_{m>0;k}  N_0^\mathrm{S}
\langle r_n'^* r_m' \rangle_\mathrm{stoch}
| \zeta^S_{ng} \rangle    \langle \zeta^S_{mk} |
\\ \nonumber &  & \mbox{ }
+
\sum_{n>0,f,h} N_n^\mathrm{S}
\langle r_n''^*  r_n'' \rangle_\mathrm{stoch}
| \zeta^S_{0h} \rangle    \langle \zeta^S_{0f} | .  
\end{eqnarray}

Equating the coefficients of
$| \zeta^S_{ng} \rangle  \langle \zeta^S_{nh} |$
yields
\begin{eqnarray}
\lefteqn{
-|\tau| [ S_n -  \delta_{n0}^\dag S_0 ] \lambda_n
} \\ \nonumber & = &
N_n^\mathrm{S}
\langle r_n^* r_n \rangle_\mathrm{stoch}
+
[1-\delta_{n0}]
N_0^\mathrm{S}
\langle  r_n'^*  r_n' \rangle_\mathrm{stoch}
\\ \nonumber &  & \mbox{ }
+ \delta_{n0}
\sum_{m>0} N_m^\mathrm{S}
\langle  r_m''^*  r_m'' \rangle_\mathrm{stoch} .
\end{eqnarray}
For the case $n=0$ this yields
\begin{equation}
-|\tau| S_0   \lambda_0
= N_0^\mathrm{S}
\langle r_0^* r_0 \rangle_\mathrm{stoch}
+
\sum_{m>0} N_m^\mathrm{S}
\langle  r_m''^*  r_m'' \rangle_\mathrm{stoch}  .
\end{equation}
Since the left hand side is positive,
and each of the terms on the right hand side is positive,
one can solve this for the drag coefficients
for any given variances of the stochastic variables
$r_0$ and $r_n''$.
For  $n>0$ one has
\begin{equation}
-|\tau| [ S_n -  S_0 ] \lambda_n
=
N_n^\mathrm{S}
\langle r_n^* r_n \rangle_\mathrm{stoch}
+
N_0^\mathrm{S}
\langle  r_n'^*  r_n' \rangle_\mathrm{stoch}.
\end{equation}
Again the left hand side is positive,
and so one can solve the above for the drag coefficients
for any given variances of the stochastic variables
$r_n$ and $r_n'$.


Equating the coefficients of
$| \zeta^S_{0h} \rangle  \langle \zeta^S_{ng} |$,
yields
\begin{eqnarray}
\lefteqn{
\frac{ -|\tau| }{2} [ S_n -  S_0 ]  \lambda_n'
} \\ \nonumber
& = &
N_0^\mathrm{S} \langle r_0^*  r_n' \rangle_\mathrm{stoch}
+  N_n^\mathrm{S} \langle r_n''^* r_n \rangle_\mathrm{stoch},
\;\; n>0 .
\end{eqnarray}
This provides the relationship between the drag coefficients
that couple the ground and excited states,
the  $\lambda_n'$,
and the correlation between the ground and excited state
stochastic coefficients,
the $\langle r_0^*  r_n' \rangle_\mathrm{stoch}$
and the $\langle r_0^*  r_n'' \rangle_\mathrm{stoch}$.

Equating the coefficients of
$| \zeta^S_{ng} \rangle  \langle \zeta^S_{0h} | $, $n>0$
yields
\begin{eqnarray}
\lefteqn{
\frac{ -|\tau| }{2} [ S_n - S_0 ]  \lambda_n'^*
} \\ \nonumber
& = &
N_n^\mathrm{S} \langle r_n^*  r_n'' \rangle_\mathrm{stoch}
+  N_0^\mathrm{S} \langle r_n'^* r_0 \rangle_\mathrm{stoch},
\;\; n>0 .
\end{eqnarray}
This is just the complex conjugate of the preceding expression.

Finally, equating the coefficients of
$| \zeta^S_{ng} \rangle    \langle \zeta^S_{mk} | $, $n \ne m$
shows that
\begin{equation}
0 =  N_0^\mathrm{S}  \langle  r_n'^*  r_m' \rangle_\mathrm{stoch}
, \;\; n \ne m .
\end{equation}

The unitary condition is also
$ \langle \hat{\cal U}(\tau)\hat{\cal U}(\tau)^\dag  \rangle_\mathrm{stoch}
= \hat{\mathrm I}$.
This is equivalent to interchanging $r_n'$ and $r_n''$
in the above.


\begin{thebibliography}{99}
\bibitem{NB1}
In  applications of quantum statistical mechanics,
the total entropy of a wave state of the sub-system
is equal to the reservoir entropy associated with that wave state,
since the internal entropy of the sub-system wave state vanishes.\cite{QSM1}
In such cases it would be best to append the subscript `r'
to the entropy to avoid any ambiguity.
\end{thebibliography}
\end{document}